\documentclass[aps,pra,amsmath,amssymb,twocolumn]{revtex4} 
\pdfoutput=1

\usepackage{graphicx}
\usepackage{dcolumn}
\usepackage{bm}
\usepackage{bbold}
\usepackage{pslatex}
\usepackage{epsfig}

\newcommand{\ket}[1]{\ensuremath{|#1\rangle}}
\newcommand{\bra}[1]{\ensuremath{\langle #1|}}

\newcommand{\be}{\begin{equation}}
\newcommand{\ee}{\end{equation}}
\newcommand{\ba}{\begin{eqnarray}}
\newcommand{\ea}{\end{eqnarray}}
\newcommand{\n}{\nonumber}

\begin{document}

\preprint{PRE/003}

\title{Rate analysis for a hybrid quantum repeater}

\author{Nadja K. Bernardes$^1$}%
 \email{nadja.bernardes@mpl.mpg.de}

\author{Ludmi{\l}a Praxmeyer$^2$}

\author{Peter van Loock$^1$}%
\email{peter.vanloock@mpl.mpg.de}

\affiliation{$^1$Optical Quantum Information Theory Group, Max Planck Institute for the Science of Light, G\"unther-Scharowsky-Str. 1/Bau 26, 91058 Erlangen, Germany}
\affiliation{Institute of Theoretical Physics I, Universit\"at Erlangen-N\"urnberg, Staudttr. 7/B2, 91058 Erlangen, Germany}
\affiliation{$^2$Institute of Physics, Nicolaus Copernicus University, ul. Grudziadzka 5, 87-100 Torun, Poland}

\date{\today}

\phantom{b}

\pacs{0000}

\begin{abstract}
We present a detailed rate analysis for a hybrid quantum repeater assuming perfect memories and using optimal probabilistic entanglement generation and deterministic swapping routines. The hybrid quantum repeater protocol is based on atomic qubit-entanglement distribution through optical coherent-state communication. An exact, analytical formula for the rates of entanglement generation in quantum repeaters is derived, including a study on the impacts of entanglement purification and multiplexing strategies. More specifically, we consider scenarios with as little purification as possible and we show that for sufficiently low local losses, such purifications are still more powerful than multiplexing. In a possible experimental scenario, our hybrid system can create near-maximally entangled ($F=0.98$) pairs over a distance of 1280 km at rates of the order of 100 Hz.
\end{abstract}

\maketitle

\section{Introduction} 

In most quantum information processes entanglement plays an essential role. It enables us not only to teleport quantum information \cite{bennett}, but also to achieve perfectly secure quantum communication \cite{ekert}. Unfortunately, the quantum channels over which entanglement is distributed are in general noisy. Owing to fundamental principles, common procedures used in classical communication, such as amplification or cloning \cite{wootters, dieks}, cannot be applied and therefore the fidelity of transmission will be limited by the length of the channel. To avoid the exponential decay with the distance and be able to perform long-distance quantum communication, quantum repeaters were proposed \cite{briegel, dur}. Instead of distributing entanglement over long distances, entanglement will be generated in smaller segments and a combination of entanglement swapping \cite{zukowski} and entanglement purification \cite{bennett, deutsch} enables one to extend the entanglement over the entire channel.

There are various promising proposals for implementing quantum repeaters. The most prominent of these approaches use some heralding mechanism based on single-photon detection to generate entangled pairs \cite{duan, childress1, childress2}. In these schemes, typically, high-fidelity entangled pairs are created, while the success probabilities in the initial entanglement distribution are very low. Other schemes, employing bright multiphoton signals, are much more efficient, but have only modest initial fidelities, since they are more sensitive to photon losses in the communication channel. As a consequence, these coherent-state-based protocols require more purification steps \cite{PvLa, Ladd, PvLb}.

Schemes for practically implementing a quantum repeater are not straightforward, not even for not too long distances such as a few hundred kilometers. The steps of entanglement distillation and swapping require advanced local quantum logic, such as two-qubit entangling gates; furthermore, the typical duration to successfully generate an entangled pair imposes severe constraints on the quantum memory decoherence times. Depending on how imperfect these operations are and how short the quantum memory decoherence time is, the final fidelity may still  exponentially decay with the total distance. Another important issue are the rates at which quantum transmission succeeds. Assuming that this rate is mainly determined by the channel transmissivity, the overall transmission rate will also decay exponentially, unless sufficient quantum memories and/or quantum error detection mechanisms are available.
	
In this paper, we will perform a study on the rates for a hybrid quantum repeater \cite{PvLa, Ladd, PvLb}. In this proposal, entanglement is created between atomic qubits through an optical coherent state. In order to simplify the analysis, we shall assume that perfect memories are available, that is, memories with infinite decoherence times. Moreover, we will assume that optimal entanglement generation probabilities and deterministic swapping routines are available. More specifically, the generalized measurements on the optical mode that lead to the initial, conditional entangled qubit pairs can be as good and efficient as allowed quantum mechanically. We will derive exact formulas for the time needed to generate an entangled pair and hence for the final rates at a given target fidelity. As opposed to previous, mainly numerical studies \cite{PvLa, Ladd, vanmeter, munro}, our rate calculations are fully analytical. For this purpose, we shall first analyze the significance of nested purifications and multiplexing with the aim of using as little as possible of these experimentally demanding techniques. Our assumption of perfect memories makes the use of probabilistic entanglement purification preferable to alternate techniques based upon deterministic quantum error correction (QEC) \cite{perseguers, jiang, fowler, munro2}. These schemes, using QEC codes, are experimentally more demanding due to their need of sufficient spatial memory resources and more complicated quantum gates for encoding and syndrome identification. In general, there will be a trade-off between the requirements on the memory decoherence times and those on the quantum error detection mechanism used to suppress the exponential fidelity decay. As mentioned, in our case, we shall focus on a scenario where memories are ideal and quantum error detection as simple as possible. The paper is structured as follows. In Sec. II, we briefly describe the hybrid quantum repeater, explaining how to generate, purify, and swap entanglement. In Sec. III, we calculate the rates to generate an entangled pair over the entire distance for a hybrid quantum repeater considering different strategies. We conclude in Sec. IV and give more details on various formulas and derivations in the appendix.

\section{Hybrid Quantum Repeater} 
\subsection{Entanglement generation}

In a hybrid quantum repeater scheme, the entanglement distribution mechanism is based on dispersive light-matter interactions, obtainable from the Jaynes-Cummings interaction Hamiltonian in the limit of large detuning \cite{schleich}. Such a Hamiltonian can be obtained by single electrons trapped in quantum dots \cite{bracker} or by neutral donor impurities in semiconductors \cite{strauf}. It leads to a conditional phase-rotation of the field mode, $\hat{U}_{int}= e^{i\theta\hat{\sigma}_{z}\hat{a}^{\dagger}\hat{a}}$, where $\theta$ is an effective interaction time, $\hat{\sigma}_{z}$ is the qubit Pauli-Z operator and $\hat{a}$ ($\hat{a}^{\dagger}$) is the annihilation (creation) operator of the electromagnetic field mode. Using a coherent state $\ket{\alpha}$ as the probe beam and an electron-spin system in a cavity (i.e., a two-level system or a ``$\Lambda$-system'' as an effective two-level system), the total output state will be ideally described as
\begin{equation}
\hat{U}_{int}\left[\frac{(\ket{0}+\ket{1})}{\sqrt{2}}\ket{\alpha}\right]=\frac{\ket{0}\ket{\alpha}+\ket{1}\ket{\alpha e^{-i \theta}}}{\sqrt{2}}.
\label{phaserotation}
\end{equation}
This interaction enables one to generate an entangled two-qubit state. First, we let a bright coherent-state pulse, or ``qubus'', which we denote as system ``B", interact with an atomic qubit superposition state, system ``A", resulting in the state (\ref{phaserotation}). The coherent state is then sent through a lossy channel and interacts in a second cavity with system ``C", resulting in an entangled state between the two qubits (systems ``A" and ``C") and the probe beam (system ``B"). This state, after local transformations, is given by \cite{PvLb}
\begin{equation}
\mu_E^2\ket{\Phi^+}\bra{\Phi^+}+(1-\mu_E^2)\ket{\Phi^-}\bra{\Phi^-}
\label{finalstate1}
\end{equation}
where
\begin{eqnarray}
\ket{\Phi^+}=\frac{1}{\sqrt{2}}\ket{\sqrt{\eta}\alpha}_B\ket{\phi^+}_{AC}+\frac{1}{2}e^{-i\eta\xi}\ket{\sqrt{\eta}\alpha e^{i\theta}}_B\ket{10}_{AC}\nonumber\\
+\frac{1}{2}e^{i\eta\xi}\ket{\sqrt{\eta}\alpha e^{-i\theta}}_B\ket{01}_{AC}, \nonumber\\
\ket{\Phi^-}=\frac{1}{\sqrt{2}}\ket{\sqrt{\eta}\alpha}_B\ket{\phi^-}_{AC}-\frac{1}{2}e^{-i\eta\xi}\ket{\sqrt{\eta}\alpha e^{i\theta}}_B\ket{10}_{AC}\nonumber\\
+\frac{1}{2}e^{i\eta\xi}\ket{\sqrt{\eta}\alpha e^{-i\theta}}_B\ket{01}_{AC}, \nonumber
\end{eqnarray}
with the maximally entangled Bell states $\ket{\phi^\pm}=(\ket{00}\pm\ket{11})/\sqrt{2}$. Photon losses in the channel are described by a beam splitter which transmits, on average, $\eta$ photons, $\xi\equiv\alpha^2 \sin{\theta}$, and $\mu_E=(1+e^{-(1-\eta)\alpha^2(1-\cos{\theta})})^{1/2}/\sqrt{2}$ ($\alpha\in\mathbb{R}$). Considering a standard telecom fiber, where photon loss is assumed to be 0.17 dB per km, the transmission parameter will be $\eta(L,L_{att})=e^{-L/L_{att}}$, where $L$ is the total distance of the channel and the attenuation length is assumed to be $L_{att}=25.5$ km.

Measuring the state of the qubus mode permits the preparation of a two-qubit entangled state. One way to achieve this final step is through homodyne detection. This is a very efficient and practical way, but the final fidelities are rather modest (e.g. $F<0.8$ for 10 km). Another slightly less practical way, but with a considerable improvement in the final fidelities, is the unambiguous state discrimination (USD) approach. High initial fidelities can then be achieved at the expense of lower entanglement generation rates. In general, using the USD measurement, fidelities can be tuned in the whole range $0.5<F<1$ for any given elementary distance $L_0$, with correspondingly smaller success probabilities for larger fidelities. In any case, the USD approach gives us ultimate bounds on the performance of the entanglement generation procedure, when the quantum mechanically optimal USD is considered \cite{PvLb}.

In the USD approach, we must be able to distinguish between the state $\ket{\sqrt{\eta}\alpha}$ and the set of states $\{\ket{\sqrt{\eta}\alpha e^{i\theta}},\ket{\sqrt{\eta}\alpha e^{-i\theta}}\}$. The fidelity between these two density operators gives a lower bound to the failure probability (the probability for obtaining an inconclusive measurement outcome),
\begin{equation}
P_{?}\geq F=\sqrt{\bra{\sqrt{\eta}\alpha}\hat{\rho}\ket{\sqrt{\eta}\alpha}},
\label{pfailure1}
\end{equation}
where
\begin{equation}
\hat{\rho}=\frac{1}{2}(\ket{\sqrt{\eta}\alpha e^{i\theta}}\bra{\sqrt{\eta}\alpha e^{i\theta}}
+\ket{\sqrt{\eta}\alpha e^{-i\theta}}\bra{\sqrt{\eta}\alpha e^{-i\theta}}).\nonumber
\end{equation}
This leads to an optimal (minimal) failure probability
\begin{equation}
P_{?}^{opt}=e^{-\eta\alpha^2(1-\cos{\theta})}.
\label{pfailure2}
\end{equation}
By looking at Eq.~(\ref{finalstate1}), it is possible to establish a connection between the fidelity of the successfully created, entangled pair (a rank-2 mixture of the $\ket{\phi^{\pm}}$ Bell states for an error-free identification of the state $\ket{\sqrt{\eta}\alpha}$) and the optimal failure probability. Considering that $\mu_E^2\equiv F$, this failure probability will be given by
\begin{equation}
P_{?}^{opt}=(2F-1)^{\eta/(1-\eta)}.
\label{pfailure3}
\end{equation}
A practical implementation of a suboptimal USD measurement based upon linear optics and photon detection can be found in \cite{PvLb}. A protocol for implementing the optimal USD is given in \cite{azuma}.

Assuming that (\ref{pfailure3}) is an optimal bound for the failure probability, the optimal upper bound for the success probability to generate an entangled pair is $P_{success}^{opt}=1-P_{?}^{opt}$. We will use this bound for the probability of success in the following sections.

\subsection{Entanglement purification and swapping}

Using the same interactions as presented in the preceding section, we are able to perform entanglement swapping and entanglement purification. In either case, local two-qubit gates are needed. Following Ref.~\cite{PvLc}, a measurement-free, deterministic controlled-phase gate can be achieved with a sequence of four conditional displacements of a coherent-state probe interacting with the two spins. More specifically, by appropriate choice of $\beta_1$ and $\beta_2$ ($\beta_1\beta_2=\pi/8$), up to a global phase and local unitaries, the total unitary operator representing the controlled-phase gate will be
\begin{equation}
\hat{D}(i\beta_2\hat{\sigma}_{z2})\hat{D}(\beta_1\hat{\sigma}_{z1})\hat{D}(-i\beta_2\hat{\sigma}_{z2})\hat{D}(-\beta_1\hat{\sigma}_{z1}).
\label{czgate}
\end{equation}
Here the operator $\hat{D}(\beta)=e^{\beta\hat{a}^{\dagger}-\beta^{*}\hat{a}}$ describes a phase-space displacement of the probe by $\beta$. In fact, it can be shown \cite{PvLc} that the sequence of (\ref{czgate}) can be achieved through uncontrolled displacements and controlled rotations of the probe via the same Jaynes-Cummings-type interaction as used for the entanglement generation. Another scheme to obtain a two-qubit gate is proposed in Refs.~\cite{PvLa,Ladd}. Although in this scheme less conditional operations are needed, making it less sensitive to losses, even without losses, a small amount of decoherence is introduced due to the remaining entanglement between the probe and the spins after the gate operation. This decoherence effect depends on $\theta$ and on the initial probe state,
and scales with $\alpha\theta^2$ (for an initial coherent-state probe with amplitude $\alpha$),
thus becoming negligible for sufficiently small $\theta$ and $\alpha\theta\sim 1$.

Single-qubit rotations, measurements, and this controlled-phase gate are sufficient resources to implement the standard purification protocol introduced in Ref.~\cite{deutsch}. From Eq.~(\ref{finalstate1}), it is possible to see that, after the USD measurement has taken place, the output state will be a rank-2 state (a mixture of two Bell states). The probability of success of the entanglement purification is
\begin{equation}
P_{purification}=F^2+(1-F)^2,
\label{ppur}
\end{equation}
and the final fidelity becomes
\begin{equation}
F_{purification}=\frac{F^2}{F^2+(1-F)^2}.
\label{fpur}
\end{equation}
As was shown in Ref.~\cite{dur}, the purification protocol introduced by Deutsch \textit{et al.} \cite{deutsch} is particularly efficient for these classes of states (i.e., it is more efficient than for full-rank mixtures).

The same kind of operations as for purification are sufficient to implement the entanglement swapping. For a state given as in Eq.~(\ref{finalstate1}), the swapping will be deterministic ($P_{swapping}\equiv1$) and its final fidelity is given by
\begin{equation}
F_{swapping}={F^2+(1-F)^2}.
\label{fswap}
\end{equation}

In a more realistic approach, considering errors in the gates caused by local losses, Eqs.~(\ref{ppur}-\ref{fswap}) will no longer be valid. According to the analysis presented in Ref.~\cite{Louis}, dissipation will be introduced in the probe mode between and during each interaction with the spins.
A coherent-state matrix element $\ket{\gamma}\bra{\beta}$ after dissipation changes to $\langle\beta|\gamma\rangle^{1-T}\ket{\gamma\sqrt{T}}\bra{\beta\sqrt{T}}$, where $T$ is the transmission parameter \cite{walls}. The interaction sequence (\ref{czgate}) will then change to terms that cause single-qubit dephasing, but also one term which causes a dephasing on both qubits; this effect scales with $(1-T)\alpha^2 \sin{\theta}^2$ (for an initial coherent-state probe with amplitude $\alpha$). For further details see Appendix A. The resulting gate is extremely sensitive to losses. However,  the aim of our work is a better understanding of the building blocks in the hybrid quantum repeater, and hence we shall minimize the effect of local losses by choosing in most parts of our analysis a sufficiently high local transmission parameter. In those figures where the parameter T is not explicitly given, it is assumed that the error is at least as small as $1-T=0.001\%$. The main influence of photon losses then occurs in the communication channel and therefore depends on the communication distance.

\section{Rate analysis}

In this section, the rates of entanglement generation over the entire distance for a hybrid quantum repeater will be calculated. The memories are considered ideal, and the entanglement connection deterministic, such that the primary source of errors will be photon losses in the channel.

\begin{figure}[h!]
\centering
\includegraphics[scale=0.42]{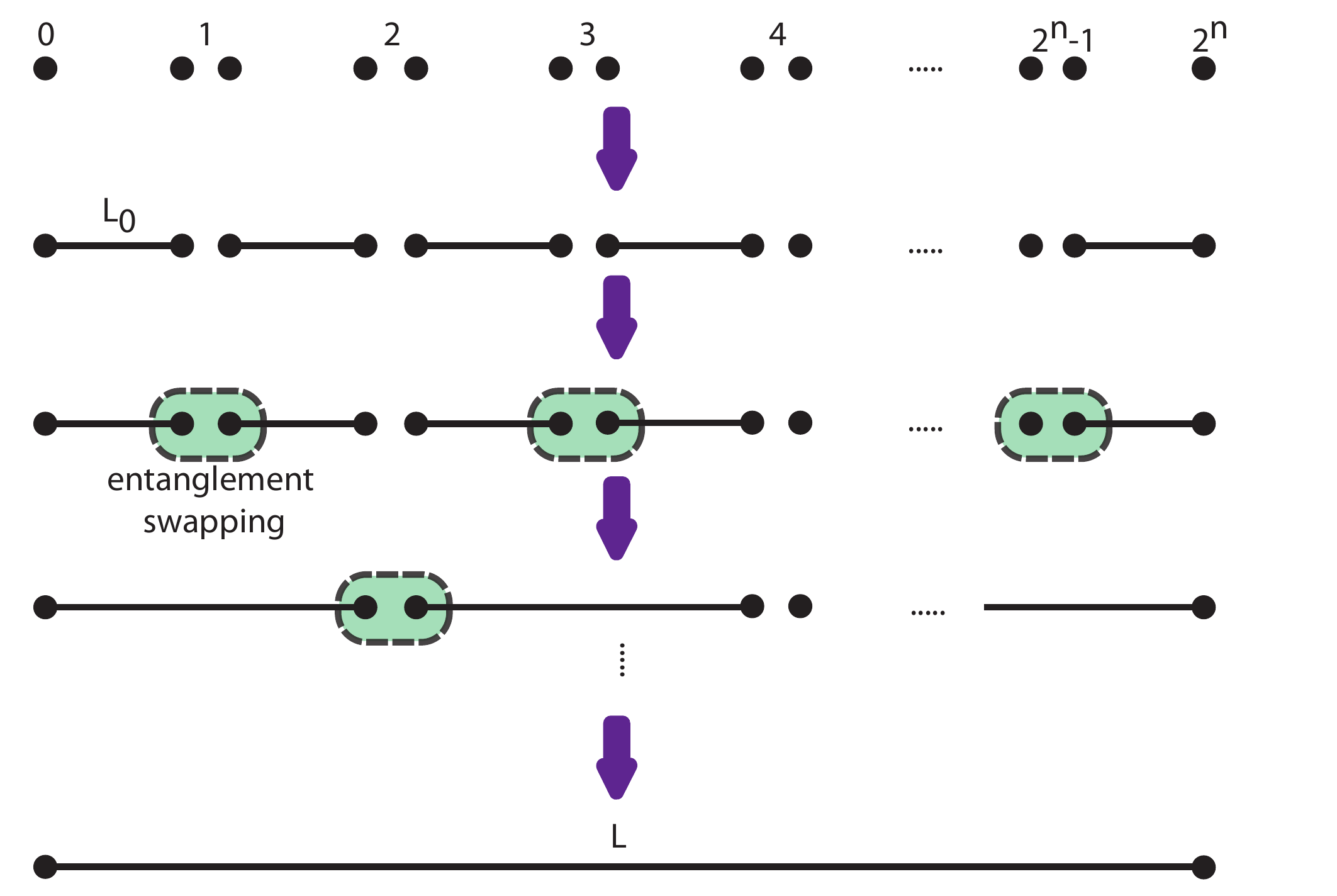}
\caption{(color online). Idealized quantum repeater. A total distance L is divided in $2^n$ segments with length $L_0=L/2^n$. Initially, entanglement is generated between neighboring repeater stations. The qubits at the intermediate stations are then connected. Finally, entanglement over the entire distance L is obtained.}
\label{qr}
\end{figure}

Let us first consider a general quantum repeater as illustrated in Fig.~(\ref{qr}). A total distance L is divided in $2^n$ segments, each of length $L_0=L/2^n$. First, entanglement is generated between the adjacent nodes, which is accomplished with probability $P_0$. Then these segments are connected, extending the entanglement from $L_0$ to $2L_0$. This step is performed many times, until the terminal nodes, separated by $L=2^nL_0$, are entangled. The initial distribution of entanglement over small segments does not prevent the final fidelity from decaying exponentially. Even if the connecting operations are perfect, we must take into account that the initial entangled pairs are not perfectly entangled pairs, and this will cause a decay of the final fidelity. As a consequence, purification and quantum memories are essential for the full quantum repeater, but their implementations introduce a series of experimental difficulties. There are various ways to optimize the combinations of swapping and purification (e.g., considering residual entanglement \cite{razavi1, razavi2}, multiplexing \cite{collins}, blind connecting measurements \cite{razavi2}). However, these approaches are more significant for the case of imperfect memories and probabilistic swappings, whereas our model uses perfect memories and deterministic swappings.

In the following sections, we will focus on the effects of {\it spatial} multiplexing compared to a parallel scheme for the hybrid quantum repeater. In a parallel repeater, the $i$th memory pair in one segment interacts only with the $i$th pair in neighboring segments, however, for the multiplexing scheme, resources can be dynamically allocated. Moreover, it will be discussed what the impact of purification on the rates is and how the two methods of multiplexing and purification compare \footnote{Note
that there are different ways to do and use multiplexing.
The more conventional application is spatial multiplexing, as typically used in discrete-variable single-photon-based
repeaters \cite{collins}. In this case, the probability to successfully generate an entangled pair, $P_0<1$, is fairly small and multiplexing means that neighboring segments even from different, parallel repeater chains
can be connected immediately after the corresponding pairs have been created.\\ 
Another type of multiplexing is temporal \cite{munro2}. In this case, the initial entanglement is generated almost instantly by transmitting in parallel sufficiently many probe pulses (after their interactions
with a corresponding number of spins) to a receiving spin that interacts with these pulses almost simultaneously
to guarantee that at least one entangled pair is created with an effective, near-unit success probability, $P_0^{\rm eff}\to 1$.
Experimentally, to make this kind of temporal multiplexing efficient, sufficiently fast local interactions and gates are needed.
As a result, when also swapping and error detection are deterministic (by replacing purification by error correction), the
full repeater protocol becomes near-deterministic, 
as opposed to our scheme which keeps both the initial entanglement generation and the entanglement purification probabilistic.\\ In principle, our rate analysis could be directly combined with the ideas of Ref.~\cite{munro2} by simply substituting our $P_0$ by an effective probability of, for instance, $P_0^{\rm eff}\equiv (1-(1-P_0)^n)$, where $n$ is the number of cavities transmitting the probe pulses
and $P_0$ in this formula is still the same function of fidelity, losses, etc., as described in the main text for our USD-based
entanglement distribution.
Then, $P_0^{\rm eff}\to 1$ can be achieved
for sufficiently large $n$ even when the original $P_0$ is as small as 10-20 \%.
However, as our goal is to keep the number of spatial resources as small and the scheme as simple as possible, we will not consider this type of temporal multiplexing here. Our setting contains perfect memories, but modest spatial resources, such that $P_0^{\rm eff}\equiv P_0$ remains of the order of 1-10 \%.}.

\subsection{Entanglement generation and swapping}

First, let us calculate the rate for generating an entangled pair for a quantum repeater in parallel without any purification or dynamical allocation of resources. For the ideal memory case, we calculate the simple case of entanglement-length doubling ($n=1$) with a single memory per half node. Since the $n$-level quantum repeater is the entanglement-length doubling of two $(n-1)$-level systems, it is essential to understand the basic process with $n=1$.

Even simpler, let us start with just one segment, $n=0$. The average time necessary to generate an entangled pair at distance $L$ (in this case $L=L_0$) is given by
\begin{equation}
\left\langle T\right\rangle_0=\frac{T_0}{P_0},
\label{Tmemoryn0}
\end{equation}
where $T_0=2L_0/c$ is the minimum time to successfully generate entanglement over $L_0$, assuming that this is the time spent on classical communication to verify the success of the entanglement generation over $L_0$ (including the initial transmission of the probe beam) and $c$ is the speed of light in an optical fiber ($2\times10^8 m/s$).

For two pairs, $n=1$, the average time necessary to generate an entangled pair at distance $L$ is then given by \cite{collins}
\begin{equation}
\left\langle T\right\rangle_1=\frac{T_0}{P_0}\frac{(3-2P_{0})}{(2-P_{0})}.
\label{Tmemory}
\end{equation}
Recall that our memories are assumed to be ideal such that one successfully created pair can be kept until a second pair is created in the neighboring segment. The result in Eq.~(\ref{Tmemory}) is equivalent to the problem of a geometrically distributed random variable with success probability $P_{0}$.

For $2^n$ pairs next to each other in $2^n$ segments, the average time needed to generate an entangled pair at distance $L$ is given by
\begin{equation}
\left\langle T\right\rangle_n= T_0 Z_n(P_0),
\label{Tmemoryn}
\end{equation}
where the average number of steps to successfully generate entanglement in all $2^n$ pairs, $Z_n(P)$, is
\begin{equation}
Z_n(P)= \sum^{2^n}_{j=1}\binom{2^n}{j}\frac{(-1)^{j+1}}{1-(1-P)^j},
\label{Zn}
\end{equation}
 and $P$ is the probability of success. Detailed calculations are presented in Appendix B.

The rate to successfully generate entanglement in all of $2^n$ pairs over L and to eventually obtain one L-distant pair can now be written as
\begin{equation}
R_n=\frac{1}{\left\langle T\right\rangle_n}=\frac{1}{T_0Z_n(P_0)}.
\label{Rn}
\end{equation}
\begin{figure}[t!]
\centering
\includegraphics[scale=0.50]{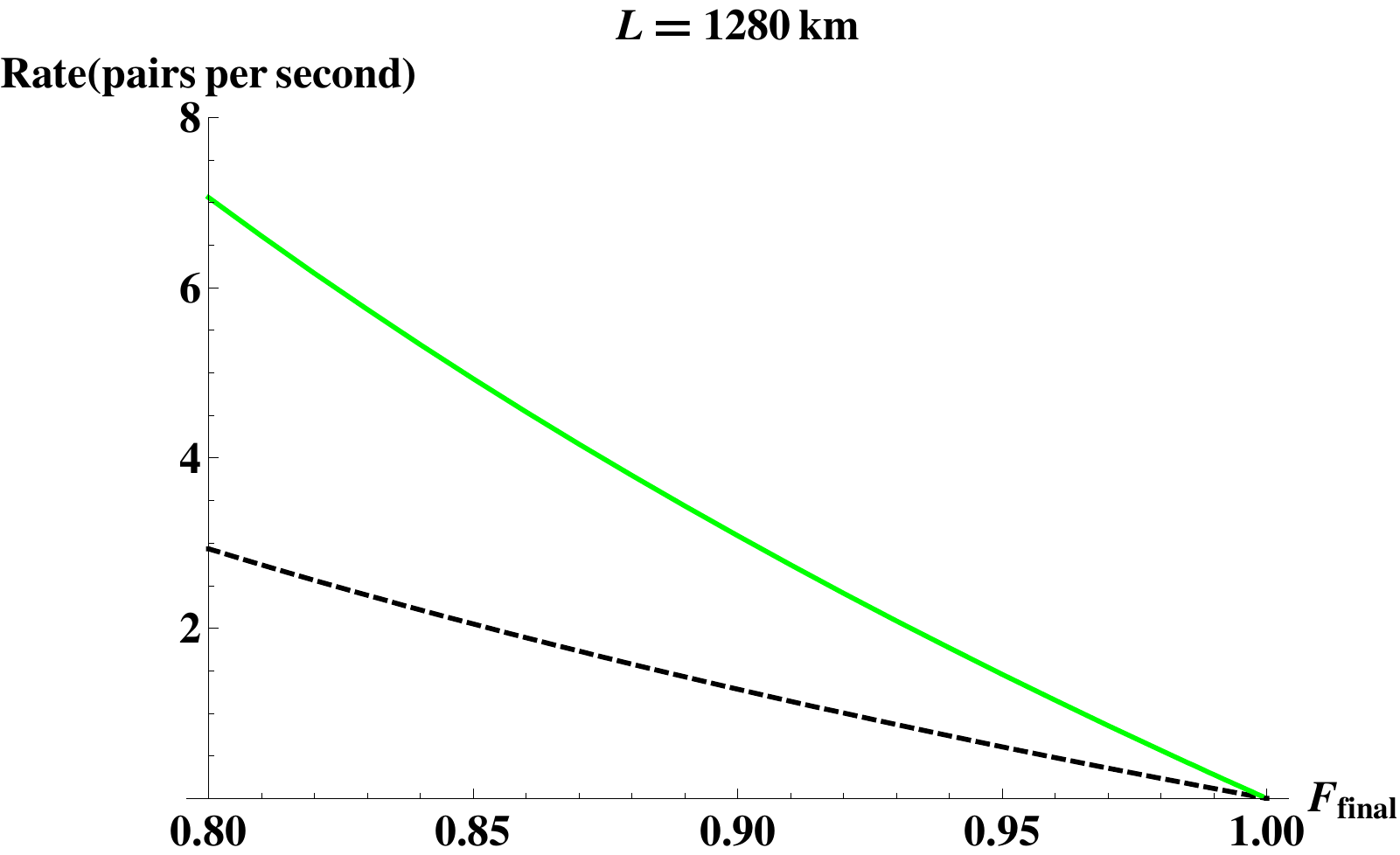}
\caption{(color online). Rates for a hybrid quantum repeater over a total distance $L=1280$ km with $L_0=20$ km without purification and without multiplexing, but including perfect memories. Comparison between the exact formula to generate entangled pairs (green line) as given by Eqs.~(\ref{Tmemoryn}-\ref{Rn}) and the approximated one (black dashed line) corresponding to $(\frac{2}{3})^n\frac{P_0}{T_0}$.}
\label{comparison}
\end{figure}
Commonly in the literature \cite{sangouard}, for small $P_0$, these rates are approximated by $\frac{P_0}{T_0 }\left(\frac{2}{3}\right)^{n}$. However, as illustrated in Fig.~(\ref{comparison}) for a total distance of $L=1280$ km and $L_0=20$ km, the approximate formula is underestimating the rates in some regimes by more than $50\%$ for our case of the USD-based hybrid repeater. 

\begin{figure}[h!]
\centering
\includegraphics[scale=0.48]{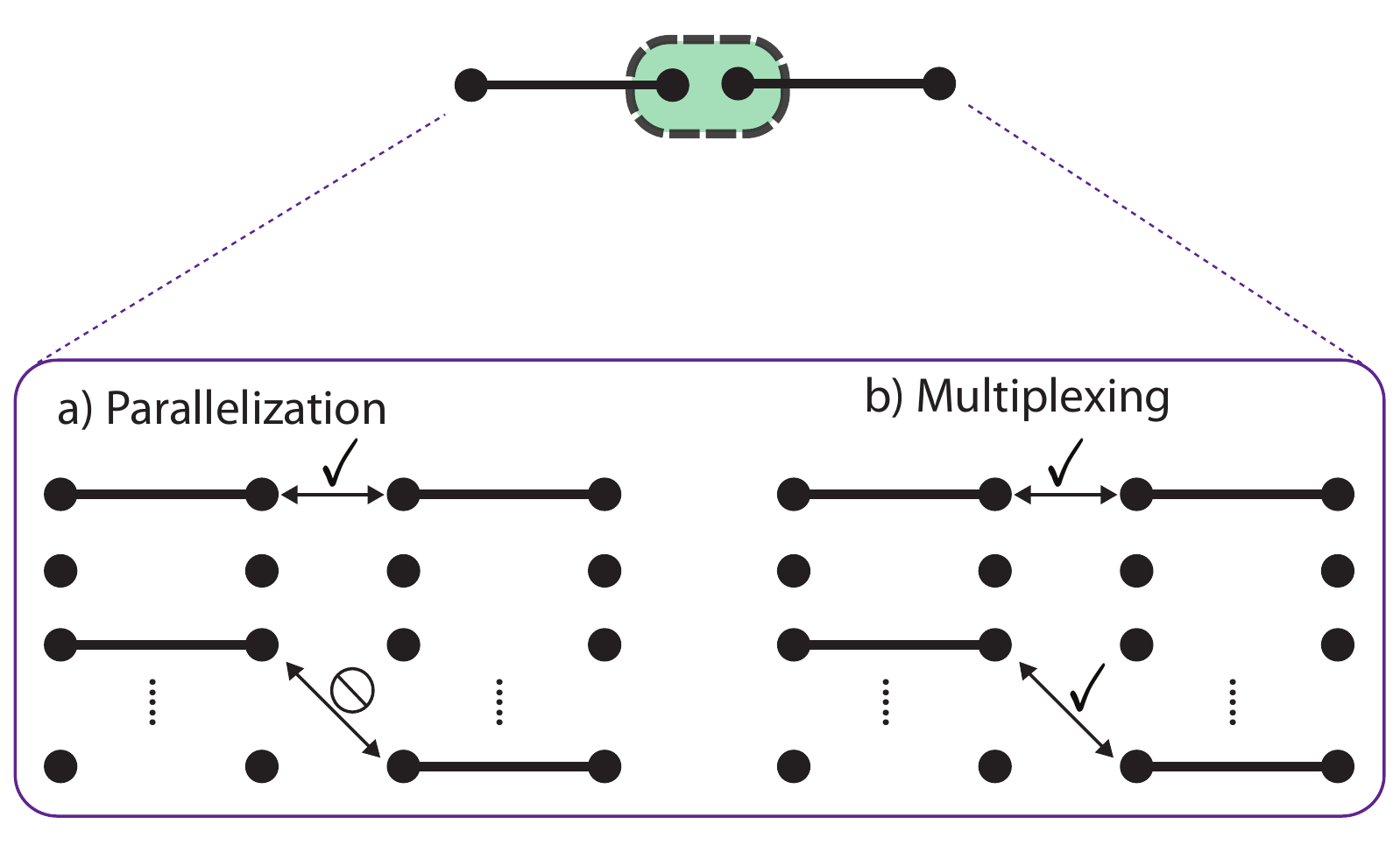}
\caption{(color online). Different strategies to connect memory elements situated in different columns. The parallel architecture a) connects only elements in the same chain. Contrary to it, multiplexing b) connects any available entangled pairs.}
\label{paralvsmult}
\end{figure}

\subsection{Multiplexing versus parallelization}

\begin{figure}[h]
\centering
\includegraphics[scale=0.50]{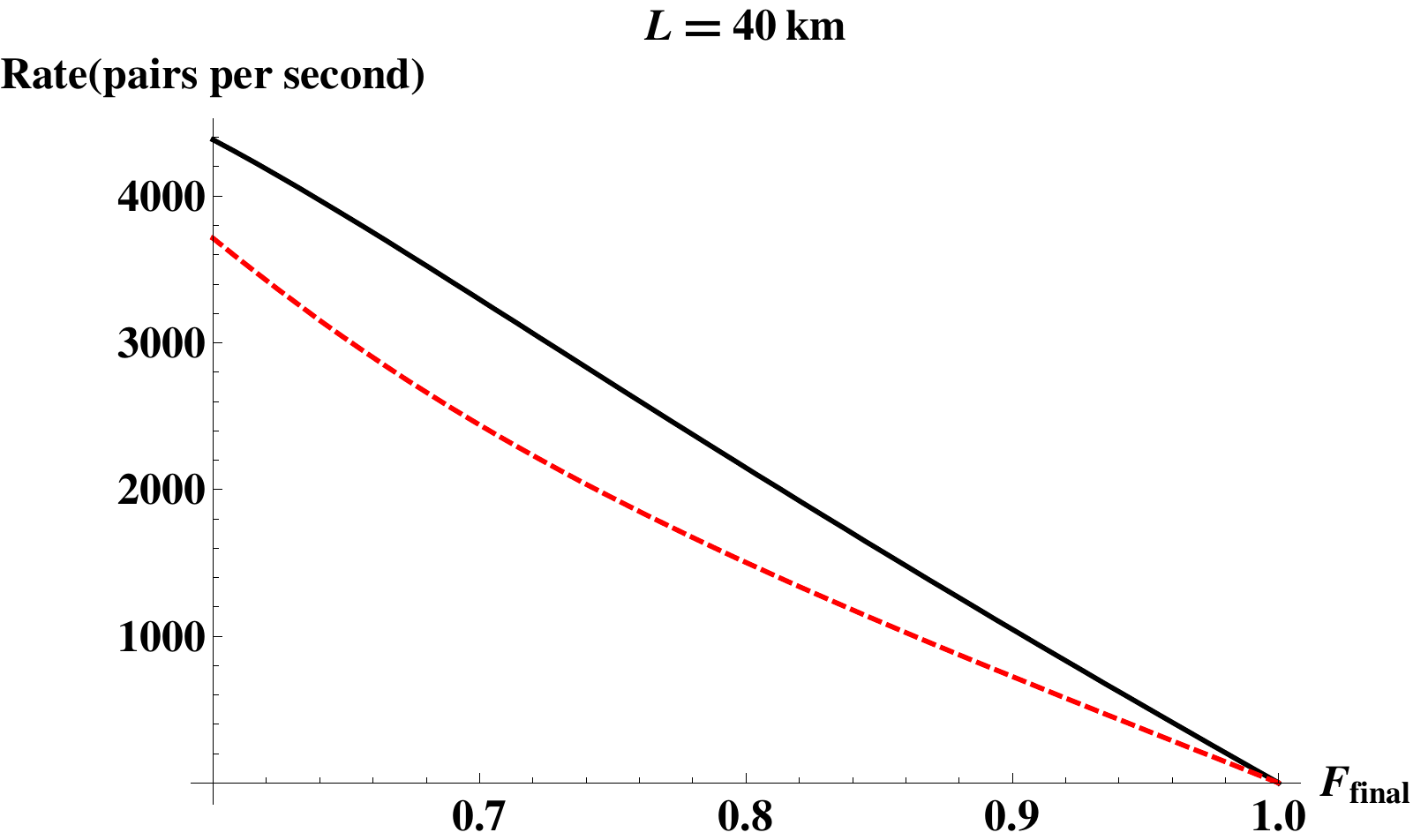}
\caption{(color online). Rates for a total distance $L= 40$ km, $L_0=20$ km. Comparison between the parallel (red, dashed line) and the multiplexed scheme (black, solid line) for 2 segments ($n=1$) and 4 memories per half node ($r=4$).}
\label{multvsparall}
\end{figure}

How would the rates be affected, if multiplexing is introduced in the system? In the multiplexed scheme,  where more than one memory per half node exist, as shown in Fig.~(\ref{paralvsmult}), entangled pairs are connected not only with pairs in the same chain (as in the parallel scheme). Instead, as soon as one entangled pair is successfully generated in one of the columns, it can be connected to a neighboring pair, no matter in which chain they are positioned. The rate to generate one entangled pair over the total distance for $n=1$ and $r$ memories per half node is given by
\be
R_{mult,1,r}=\frac{1}{T_0Z_{mult,1,r}(P_0)}=\frac{1}{T_0}\frac{1-(1-P_0)^{2r}}{1+2(1-P_0)^r}.
\label{ratemult}
\ee
For further details see Appendix C. Plotting this rate for a total distance of $L=40$ km and four memory pairs per segment, $r=4$, see Fig.~(\ref{multvsparall}), it is possible to confirm that the multiplexed scheme performs better than the parallel one. However, an experimental implementation of this type of spatial multiplexing would require fairly demanding feedforward techniques on the pulses that enact the gates for entanglement swapping. It is therefore more practical and efficient to operate the repeater in parallel, provided perfect memories are available, as it is the case in our scenario. In the following sections, the influence of purification will be studied. By comparing purification with multiplexing, we will justify why we shall not consider multiplexing in the further analysis.

\subsection{Entanglement purification}

Taking into account that purification is an expensive task, either in terms of spatial or in terms of temporal resources, the most efficient and simplest extension beyond those schemes described in the preceding sections would employ just one round of purification at the first nesting level. Intuitively, considering the purification procedure probabilistic and entanglement swapping deterministic, performing the purification at the beginning will make better use of the memories than by doing it at the end or somewhere in between; for a more quantitative justification, see Appendix D. Starting again with $n=0$, already initially there is a need to generate at least two pairs, such that purification is possible. In this way, the average time needed to generate a purified pair will be
\begin{equation}
\left\langle T\right\rangle_{purif,0}= \frac{T_0}{P_0P_1}\frac{(3-2P_{0})}{(2-P_{0})}.
\label{Tmemory1purif}
\end{equation}
Here we employ the same purifying protocol as the one introduced in \cite{zeilinger}; in this case the probability of one round of purification to succeed, $P_1$, is given by Eq.~(\ref{ppur}).

For $n>0$, the calculation of the rates is not straightforward. In this regime, we have found an upper and a lower bound for the time needed to generate $2^n$ purified pairs. The former times, which give us the lower bounds for the rates, correspond to those cases when purification will start only after all the pairs are successfully generated, even if two pairs in the same column are already present after a shorter time. In this slowest case, the average time needed to generate $2^n$ one-round purified pairs is given by
\begin{equation}
\left\langle T\right\rangle_{purif,upper,n}= T_0 Z_{n+1}(P_0)Z_n(P_1).
\label{Tmemorynpurifupper}
\end{equation}
For the most optimistic case, corresponding to the fastest possible way to achieve purification, we imagine that if purification fails, it is not necessary to start from the beginning, trying to generate new pairs again, but we assume that the two pairs necessary for purification are still available, imagining that they have not been destroyed. The average time needed to generate $2^n$ one-round purified pairs is then given by
\begin{equation}
\left\langle T\right\rangle_{purif,lower,n}= T_0(Z_{n+1}(P_0)+Z_n(P_1)).
\label{Tmemorynpuriflower}
\end{equation}

Aiming to find a compromise between the upper and lower bounds for the rates, the rates can be calculated in an approximate fashion, combining the ideas from the recurrence and the exact formula. The average time needed to generate one purified pair will be
\begin{equation}
\left\langle T\right\rangle_{purif,approx,0}= \frac{T_0}{P_{L_0}},
\label{Tmemory1purifapprox}
\end{equation}
where $P_{L_0}$ is an effective probability, from (\ref{Tmemory1purif}), $P_{L_0}={P_0P_1}\frac{(2-P_{0})}{(3-2P_{0})}$. From this, it follows that
\begin{equation}
\left\langle T\right\rangle_{purif,approx,n}= T_0 Z_{n}(P_{L_0}).
\label{Tmemorynpurifapprox}
\end{equation}

\begin{figure}[h!]
\centering
\includegraphics[scale=0.50]{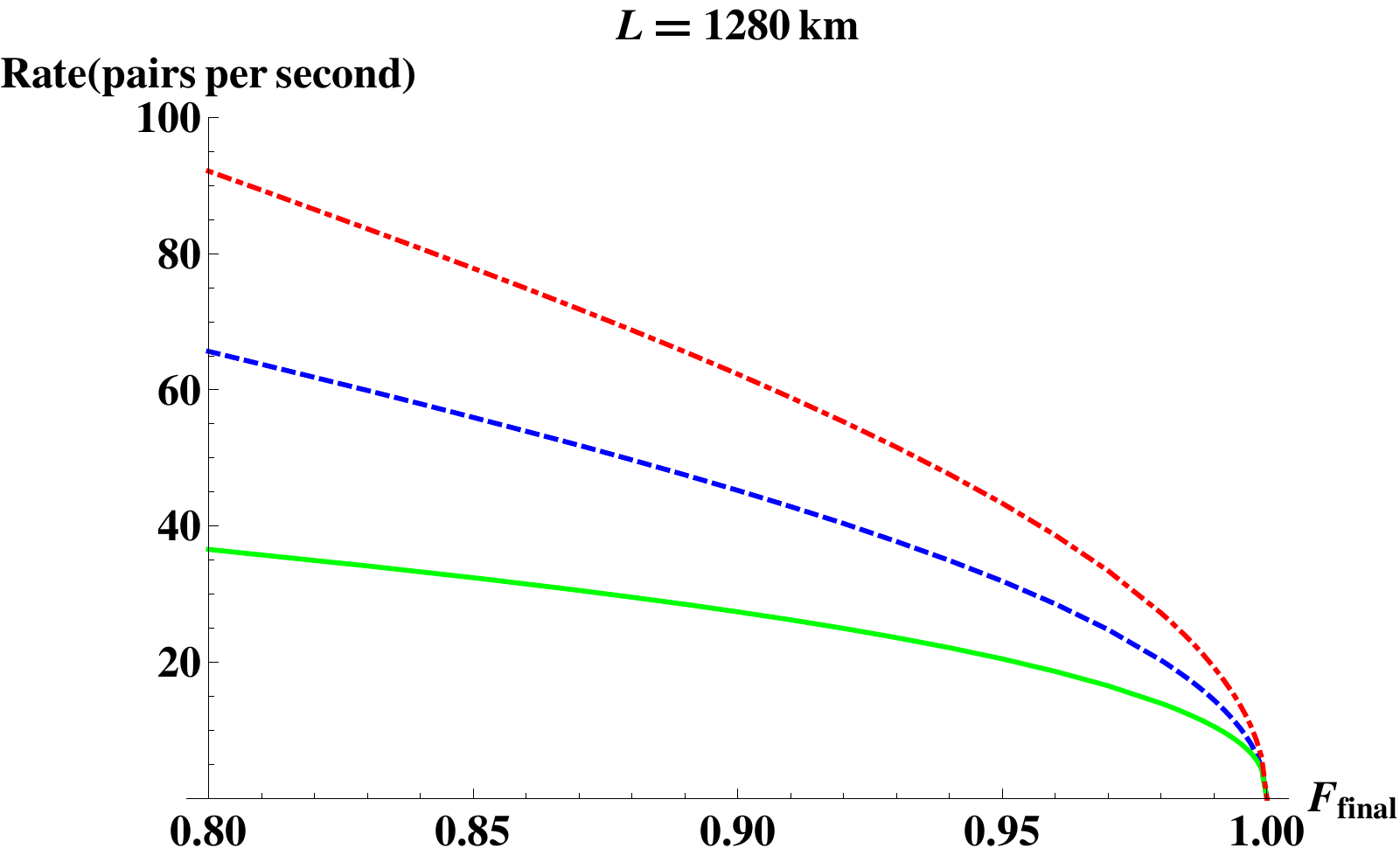}
\caption{(color online). Rates for a hybrid quantum  repeater over a total distance $L= 1280$ km and $L_0=20$ km. Comparison between the lower bound (green, solid line), upper bound (red, dot dashed line) and the approximate formula (blue, dashed line). }
\label{ratesupperlowapprox1280}
\end{figure}

How are the upper bound, the lower bound, and the approximate times, and their corresponding rates, related to each other? Fig.~(\ref{ratesupperlowapprox1280}) shows for a total distance $L=1280$ km, the rates in these three cases. The approximate formula gives a result which always stays in the middle between the upper and the lower bounds. The rates are then given by
\begin{eqnarray}
\label{ratenpurif}
R_{purif,n}&=&\frac{1}{\left\langle T\right\rangle_{purif,approx,n}}\\
&=&\frac{1}{T_0 Z_n(P_{L_0})}.\nonumber
\end{eqnarray}


\begin{figure}[h!]
\centering
\includegraphics[scale=0.50]{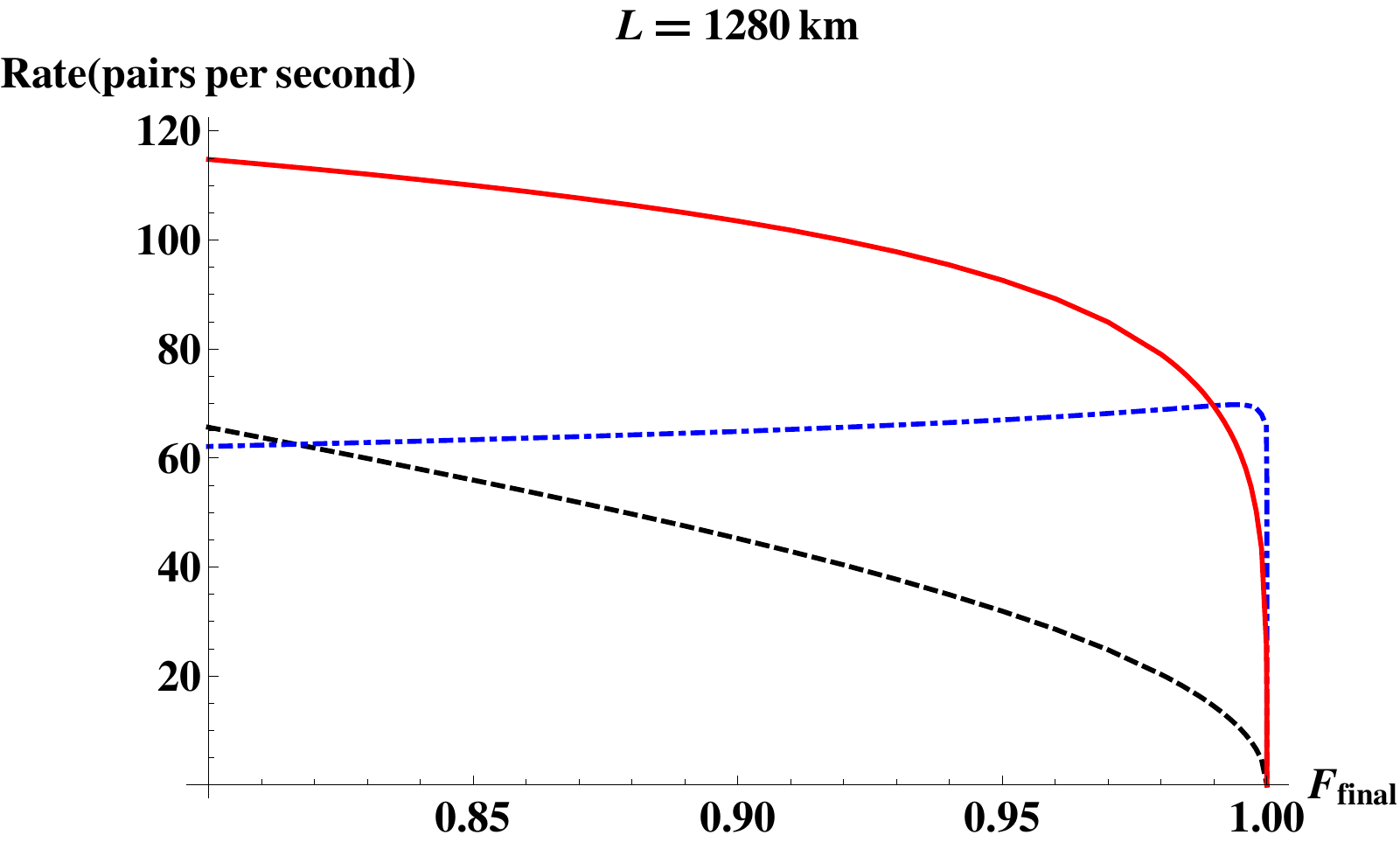}\\ \qquad \\
\includegraphics[scale=0.50]{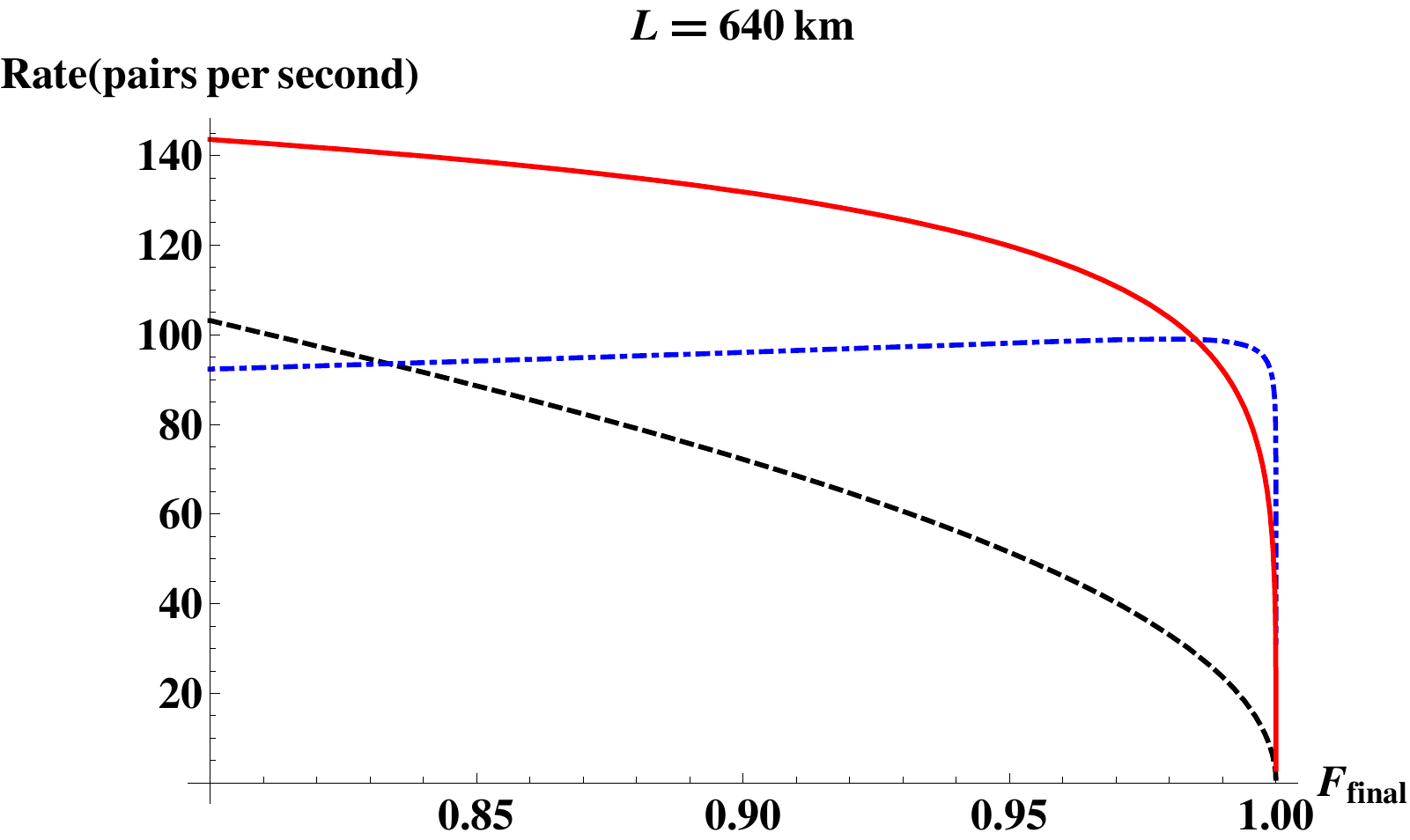}\\ \qquad \\
\includegraphics[scale=0.50]{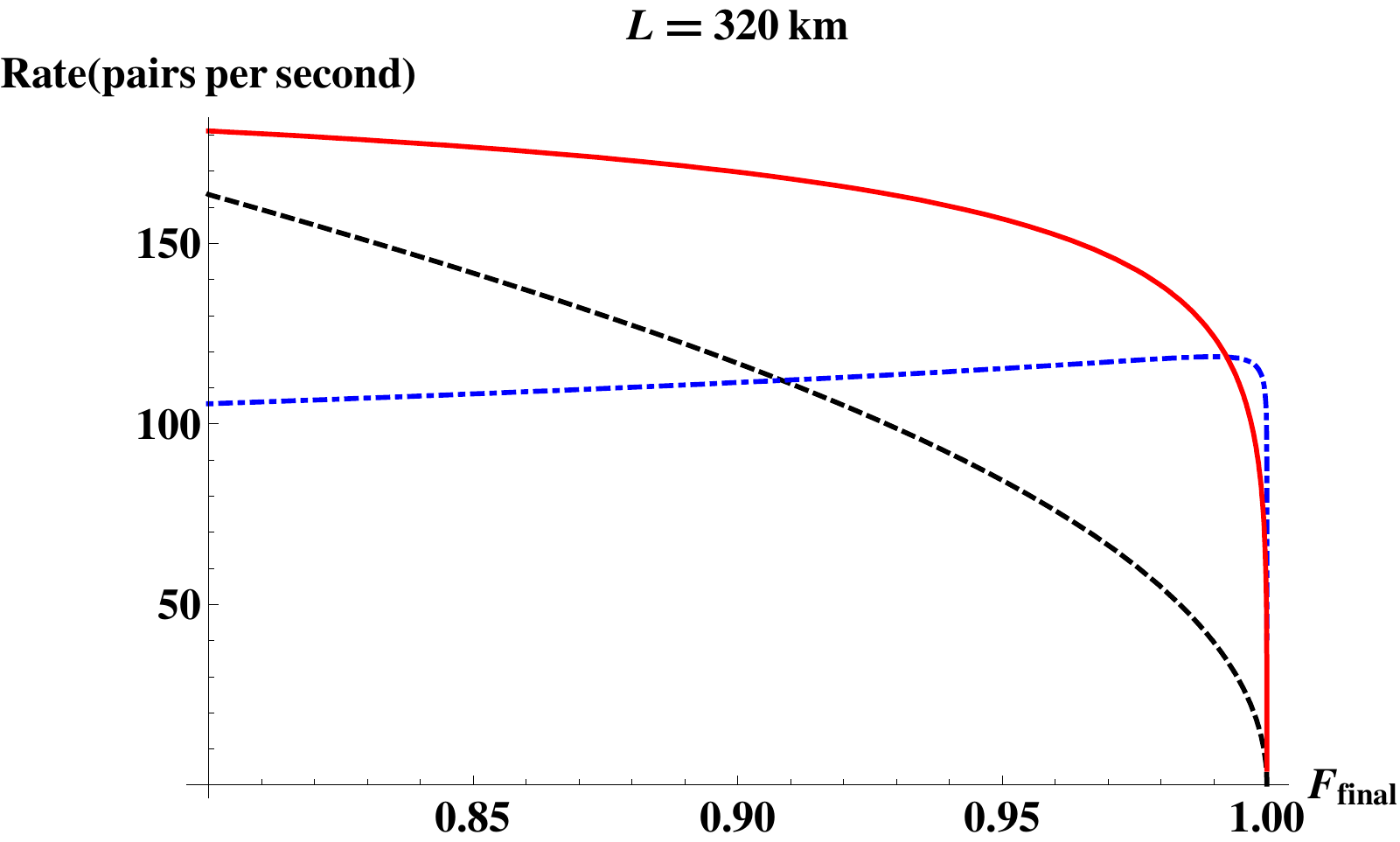}
\caption{(color online). Rates for total distance $L= 320$ km (below), $L= 640$ km (center), $L= 1280$ km (above) with $L_0= 20$ km. Comparison between the rates for a hybrid quantum repeater with one round of purification (black, dashed line), two rounds of purification (red, solid line), and three rounds of purification (blue, dot dashed line) at the first nesting level.}
\label{ratemanyroundspurif}
\end{figure}

There is one remaining question: do more rounds of purification at the first nesting level always increase the rates for the hybrid quantum repeater? Utilizing (\ref{ratenpurif}) we plotted in Fig.~(\ref{ratemanyroundspurif}) the rates to generate an entangled purified pair over different total distances applying one, two, or three rounds of purification \footnote{$P_{L_0}$ varies depending on the number of rounds of purification. For one round, $P_{L_0,1purif}={P_0P_1}\frac{(2-P_{0})}{(3-2P_{0})}$. For two rounds, $P_{L_0,2purif}={P_{L_0,1purif}P_1}\frac{(2-P_{L_0,1purif})}{(3-2P_{L_0,1purif})}$. And for three rounds, $P_{L_0,3purif}={P_{L_0,2purif}P_1}\frac{(2-P_{L_0,2purif})}{(3-2P_{L_0,2purif})}$.}. 
We see that not always does the increase of rounds of purification result in an increase of the rates. More specifically, for this scheme and these distances, three rounds of purification only increase the rates for very high final fidelities. It should be pointed out here that the number of initial resources \footnote{These initial resources can be either spatial or temporal, where in the latter case,
for every new round of purification, a fresh initial pair is created (so-called entanglement pumping \cite{dur}). Such an approach
requires a minimum of spatial resources \cite{childress1,childress2}, however, a maximum of time. This means one needs extremely good memories and,
nonetheless, the total rates remain fairly low. Even though in our setting we do assume perfect memories,
throughout we shall stick to the faster standard purification methods at the first nesting level
using a sufficient initial supply of spatial resources, similar to Ref.~\cite{PvLa}.}
increases also with the number of rounds of purification, and so, in a more appropriate analysis, keeping the numbers of initial resources the same, the rates for multiple-rounds of purification should even perform worse.

\subsection{Multiplexing versus purification}
\begin{figure}[h!]
\centering
\includegraphics[scale=0.50]{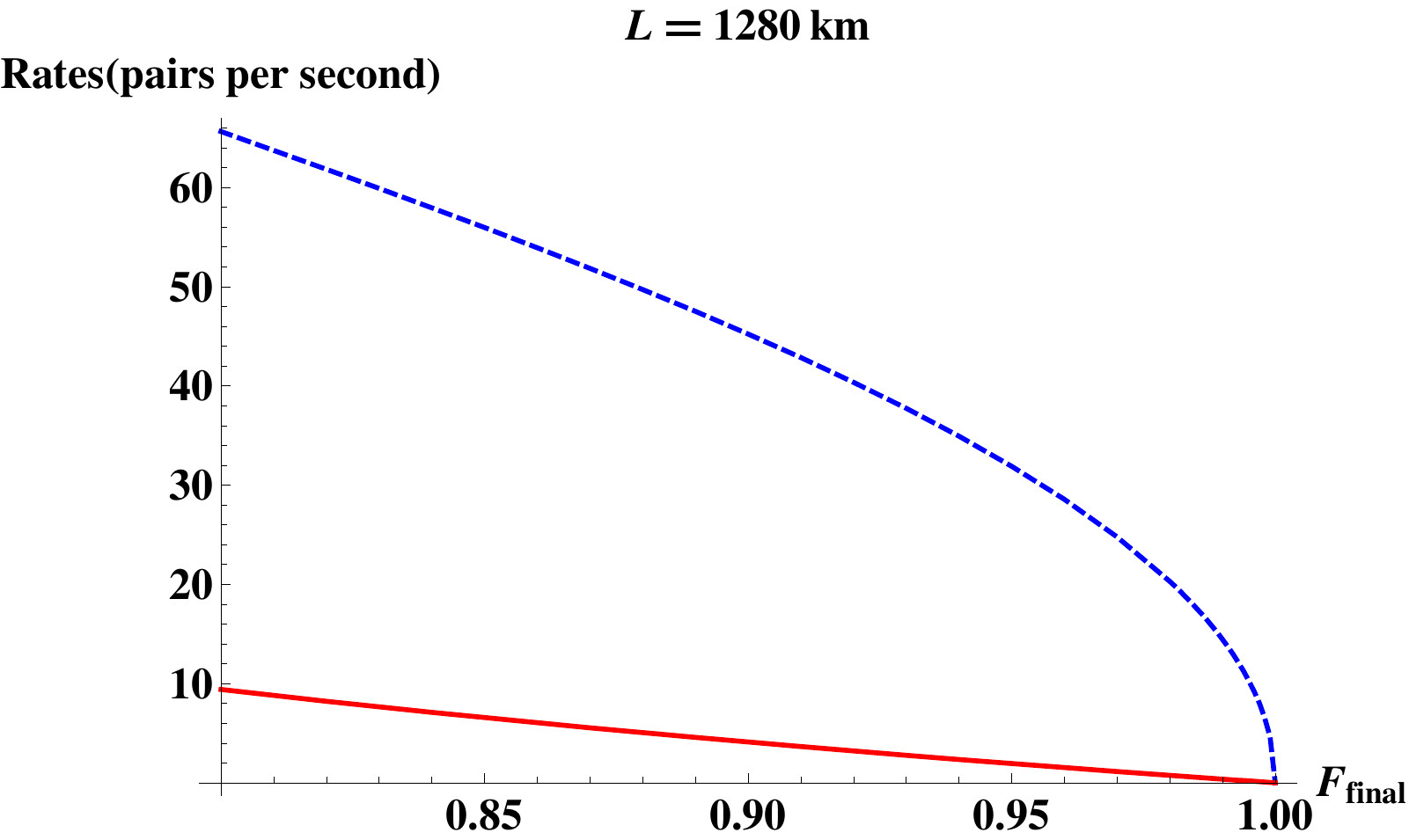}\\ \qquad \\
\includegraphics[scale=0.50]{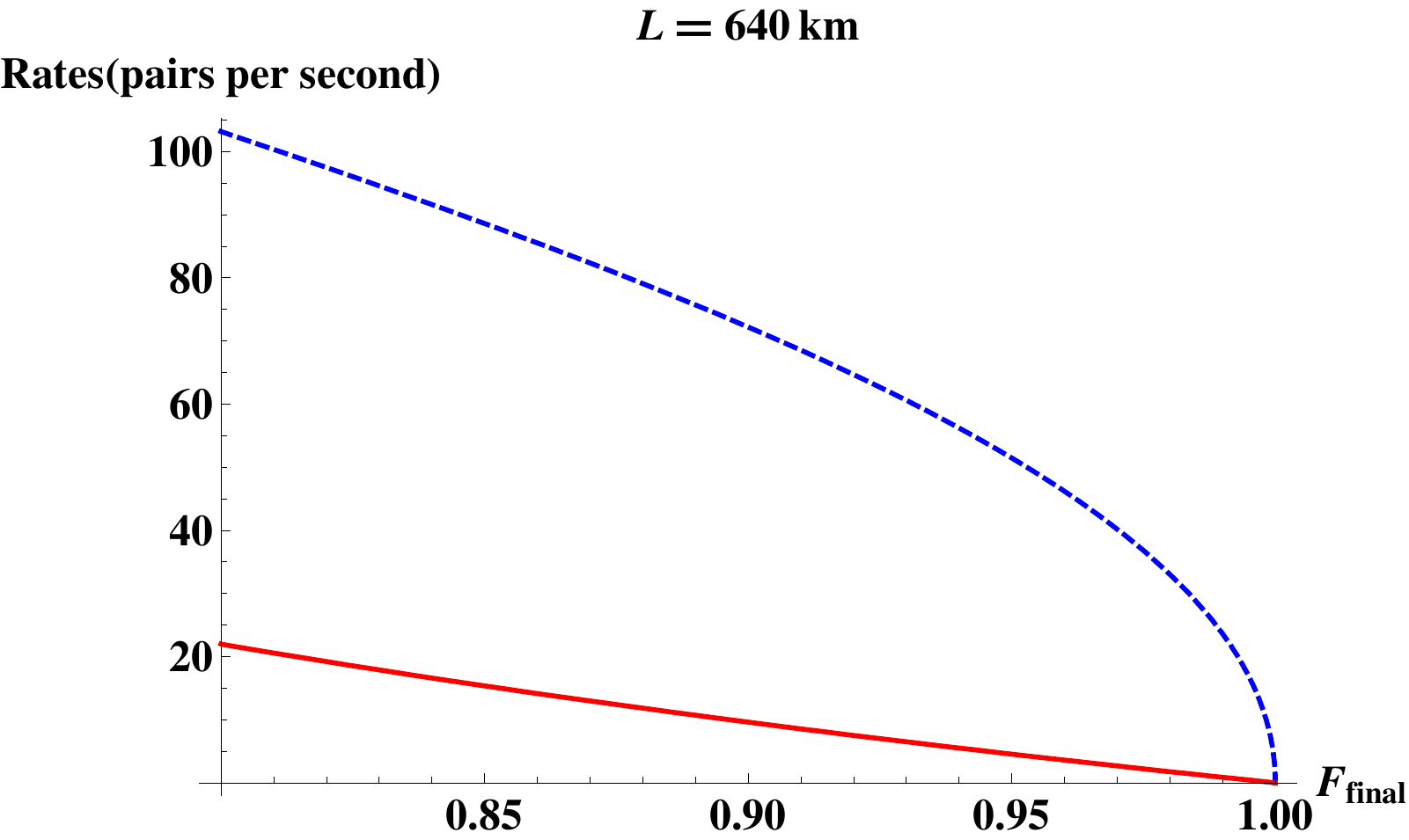}\\ \qquad \\
\includegraphics[scale=0.50]{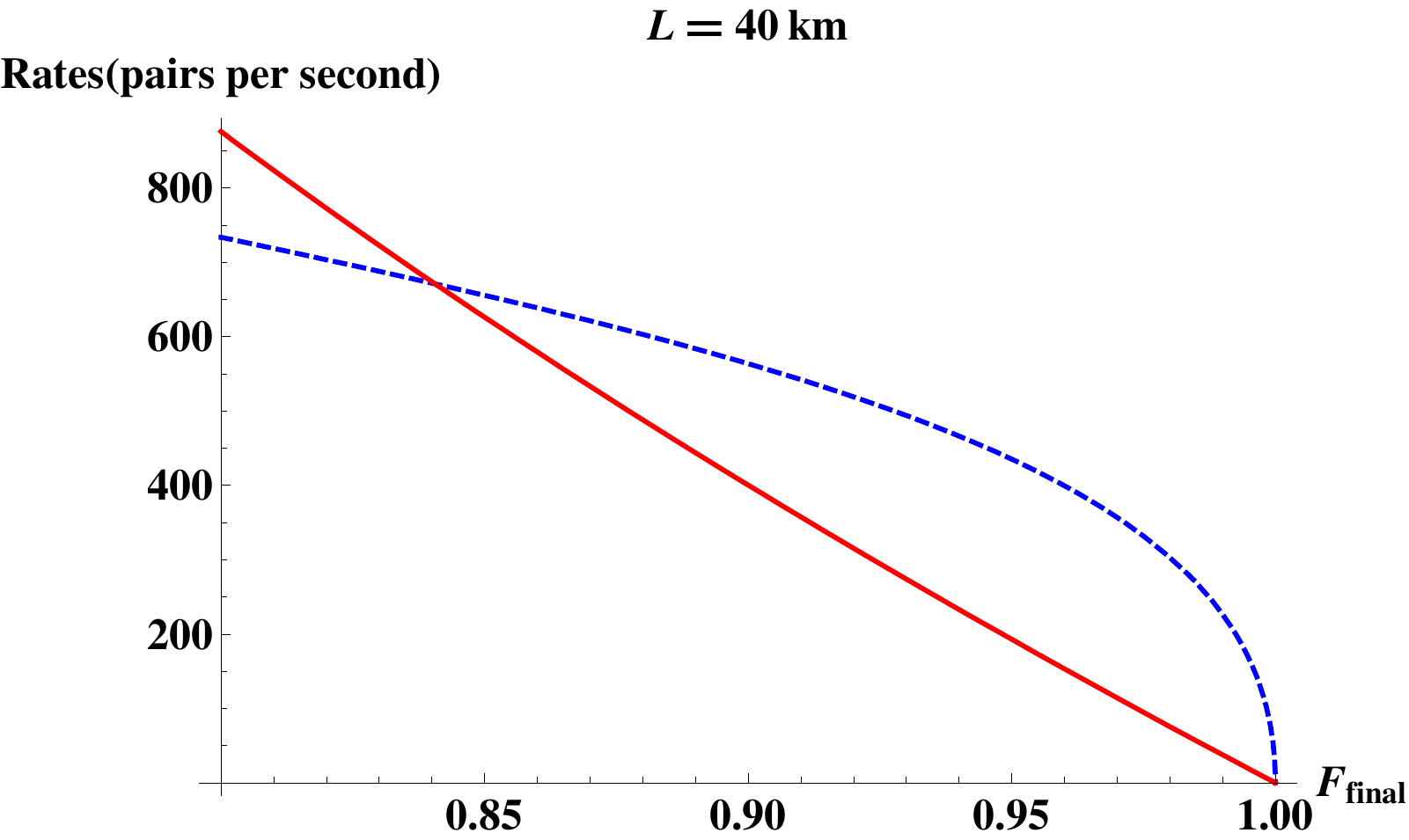}
\caption{(color online). Rates to generate one entangled pair over a total distance of $L= 40$ km (below), $L= 640$ km (center) and $L= 1280$ km (above) with $L_0= 20$ km. Comparison between multiplexing for 2 memories per half node (red solid line) and one round of purification in the first nesting level (blue dashed line). }
\label{ratesmultvspurif}
\end{figure}

After analyzing the effects of multiplexing and purification independently, we would like to compare both strategies. Although in Eq.~(\ref{ratemult}) the rates for $n=1$ and $r$ memories per half node have been calculated, for $n>1$, there is not such an analytical formula. In this case, the rates were calculated through an effective probability according to Eq.~(\ref{ratenpurif}) with different $P_{L_0}$ in each case. For the scheme with purification, we have $P_{L_0}=P_0P_1\frac{(2-P_0)}{(3-2P_0)}$ and for multiplexing, $P_{L_0}=\frac{1-(1-P_0)^{2r}}{1+2(1-P_0)^r}$. We kept the number of initial resources equal in both schemes (two memories per half node). As it is possible to observe in Fig.~(\ref{ratesmultvspurif}), even for a total distance as small as $L=40$ km, for sufficiently high fidelities, the scheme with just one round of purification is providing higher rates. For longer distances, $L=640$ km and $L=1280$ km, one round of purification is clearly more powerful than multiplexing without purification.

\begin{figure}[h!]
\centering
\includegraphics[scale=0.50]{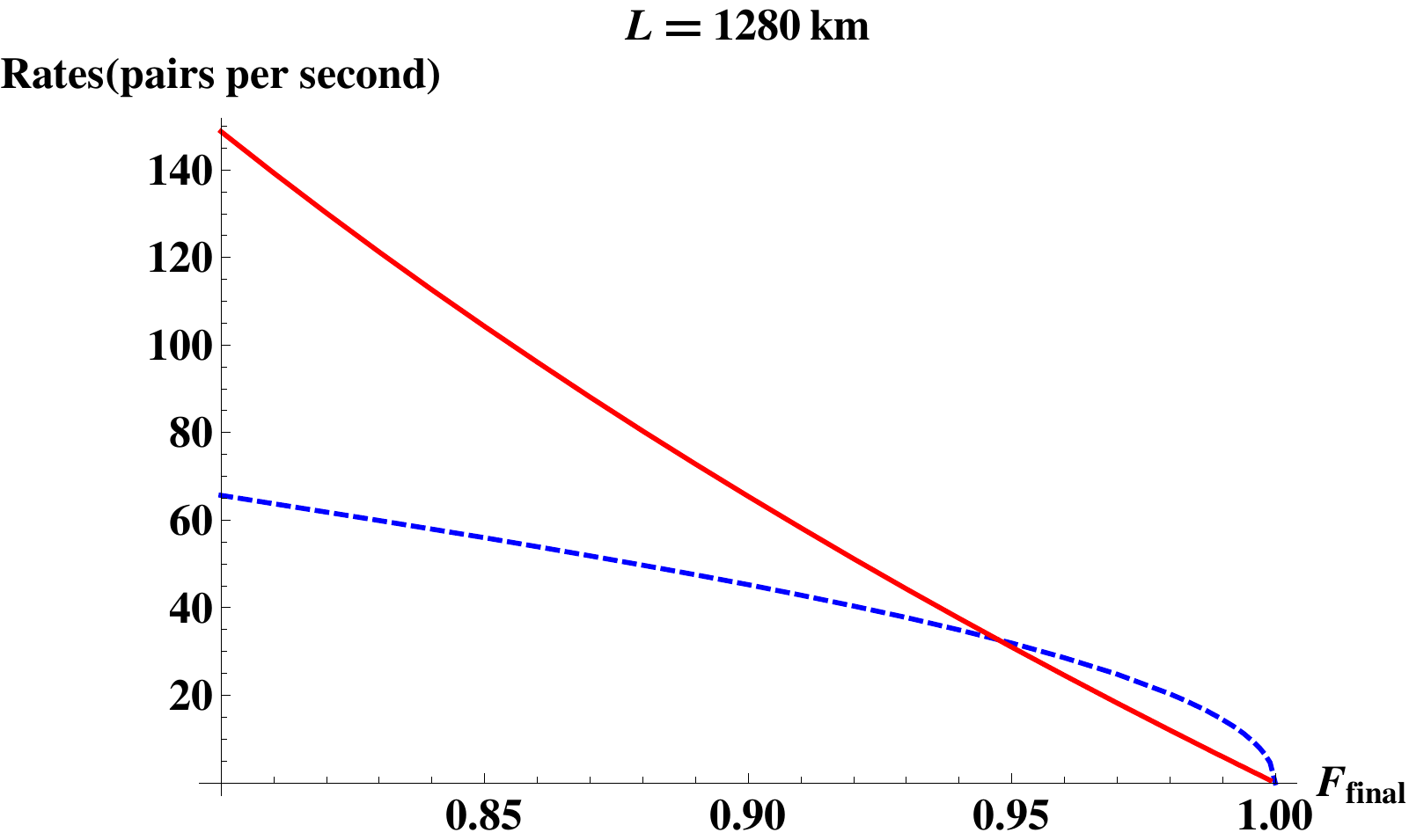}\\ \qquad \\
\includegraphics[scale=0.50]{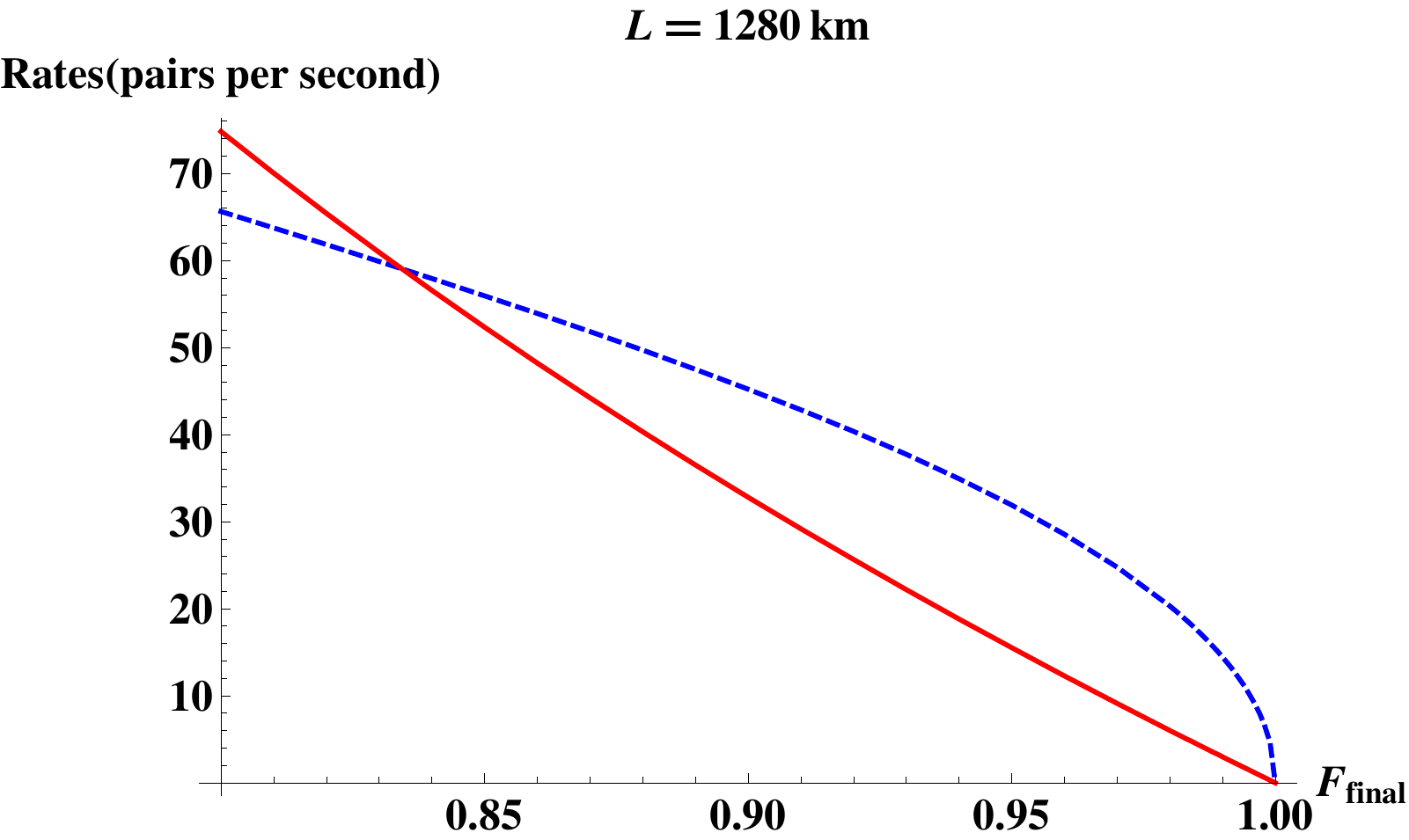}
\caption{(color online). Rates to generate one entangled pair for a total distance of $L= 1280$ km with $L_0= 20$ km. Comparison between one round of purification in the first nesting level (blue dashed line) and multiplexing (red solid line) for 32 memories per half node (above) and for 16 elements memories per half node (below). }
\label{ratesmultvspurifmoreelements}
\end{figure}

One may argue that the effects of multiplexing become more significant in a configuration with more memories per half node, i.e., bigger $r$. However, as can be seen from Fig.~(\ref{ratesmultvspurifmoreelements}), the rate to generate an entangled pair after one round of purification (only two memories per half node) is higher than the rate to generate an entangled pair for a scheme using multiplexing with 32 available memory pairs per segment, $r=32$, for fidelities larger than $~0.95$. For multiplexing schemes with $r=16$,  the purification is performing better already for a final fidelity $F_{final}>0.84$. Taking into account all these results and the fact that multiplexing is difficult to implement, multiplexing will not be considered in our final analysis. It has to be pointed out here that this argument is not valid anymore if the losses considered in the controlled- phase gates are bigger than $0.01\%$ (see Sec.~II.B and App.~A).

\subsection{Results}

\begin{figure}[h]
\centering
\includegraphics[scale=0.50]{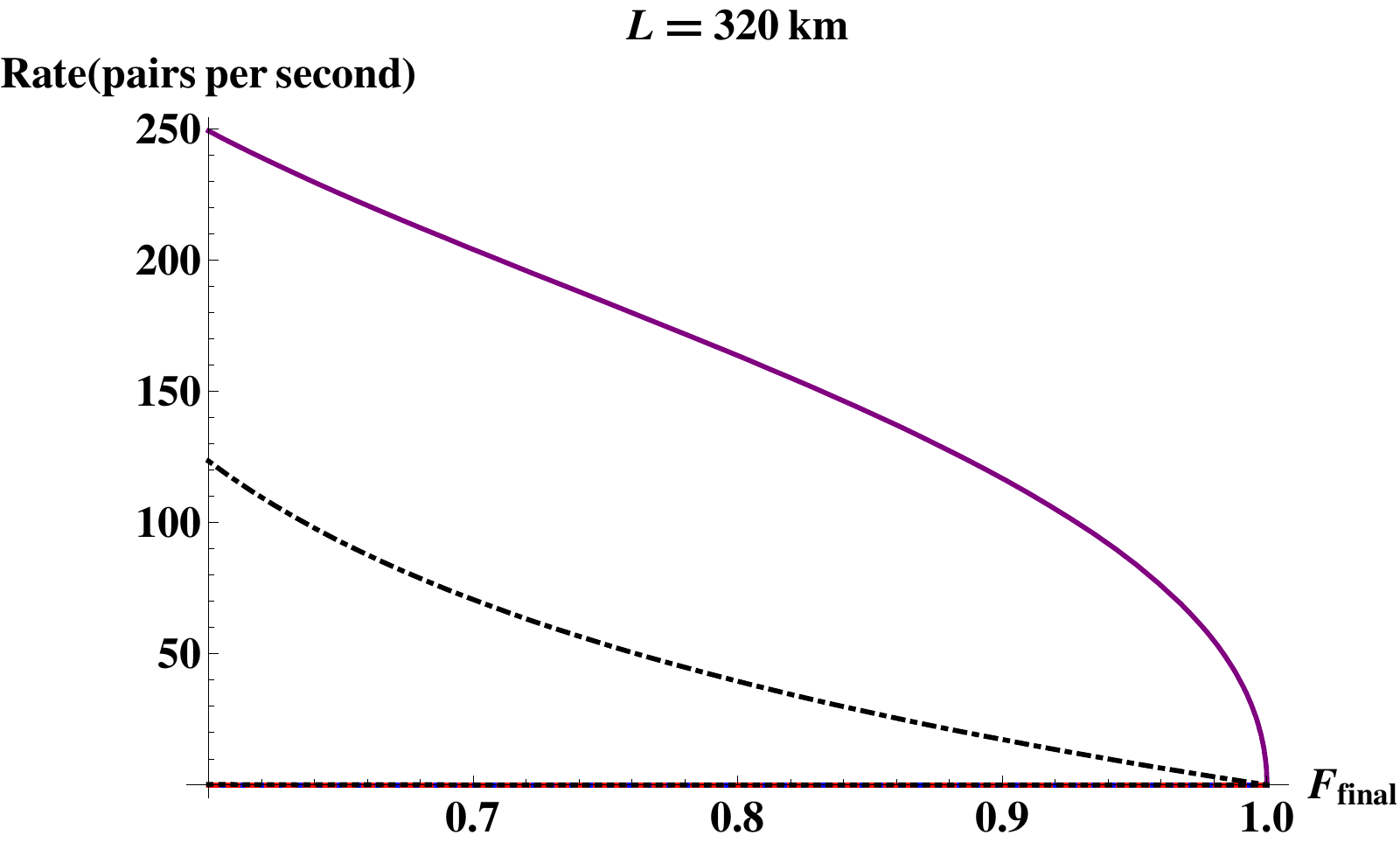}\\ \qquad \\
\includegraphics[scale=0.50]{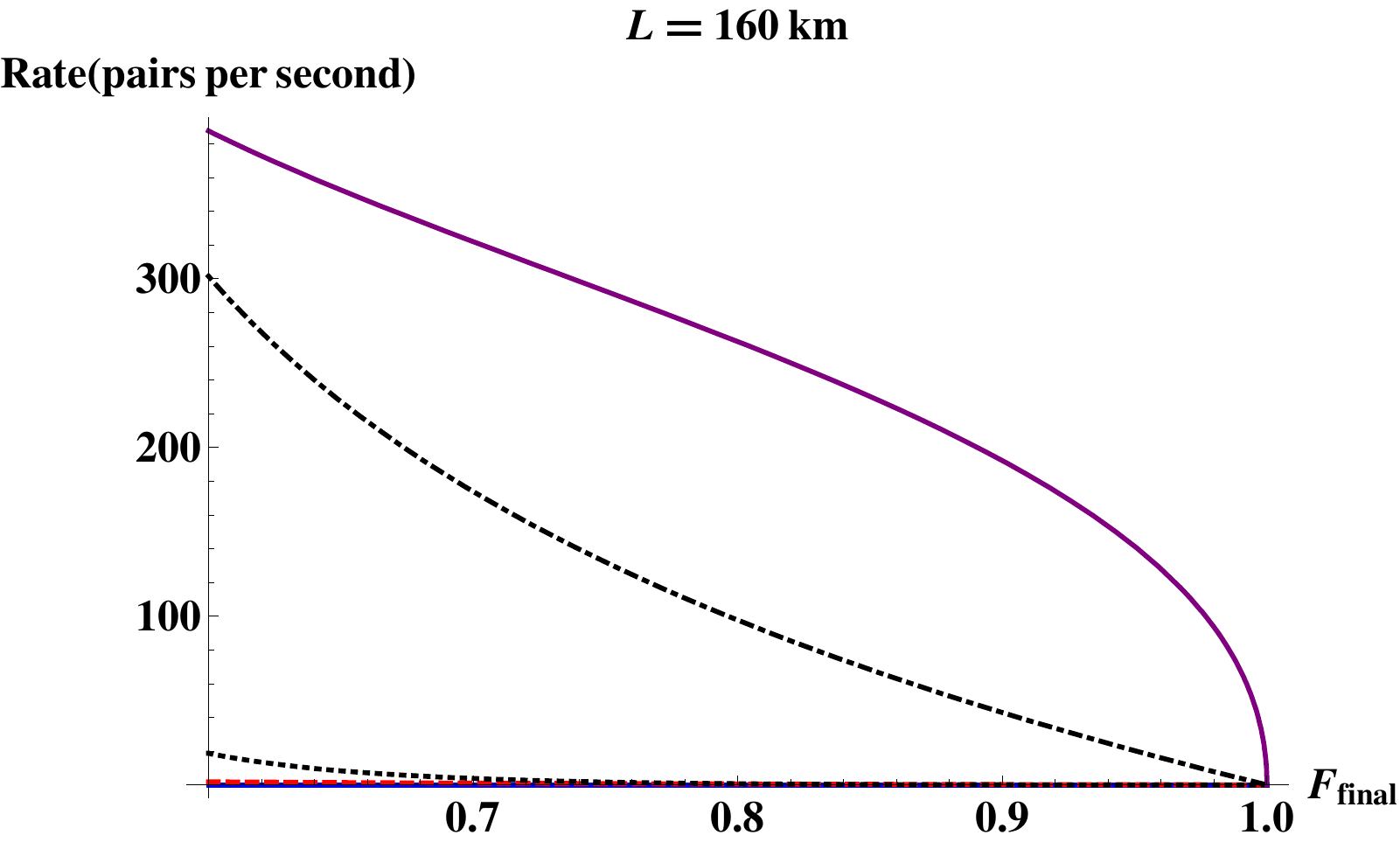}\\ \qquad \\
\includegraphics[scale=0.50]{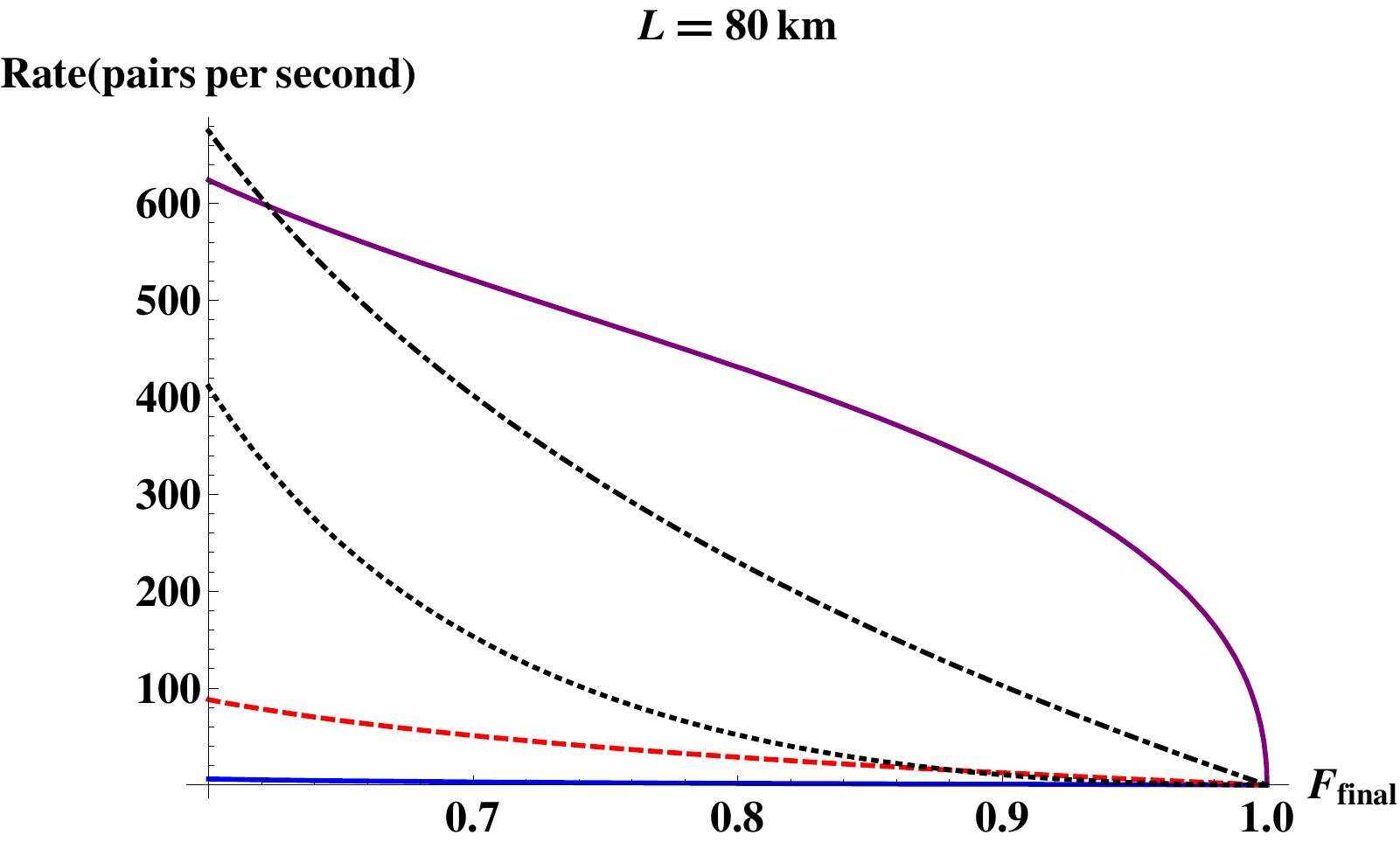}
\caption{(color online). Rates for total distance $L= 80$ km (below), $L= 160$ km (center), $L= 320$ km (above) with $L_0= 20$ km. Comparison between direct transmission with one round of purification (blue, thick line), direct transmission without purification (red, dashed line), quantum relay (black, dotted line), hybrid quantum repeater with one round of purification (purple, thick line), and hybrid quantum repeater without purification (black, dot dashed line). }
\label{ratesdifschemes}
\end{figure}

In Fig.~(\ref{ratesdifschemes}) the rates for total distances $L= 80$ km, $160$ km, $320$ km are shown, with $L_0$ always equal to $20$ km. We have compared the case where one round of purification takes place at the beginning before the connecting steps (solid lines) to the case with no purification (dashed/dotted lines). We have further compared the rates for the hybrid quantum repeater to the rate upper bounds for transmitting the state directly (i.e., without swapping) and to the rates of a quantum relay scheme, where all entangled pairs have to be successfully generated in each segment at the same time (i.e., without using memories). Even for a relatively small total distance, $L= 80$ km, the hybrid quantum repeater is performing better than the quantum relay or direct transmission. For larger distances,  $L=160$ km and $320$ km, the difference between the rates of those schemes is even bigger, and the rates for the quantum relay and direct transmission are effectively zero.

\begin{figure}[h!]
\centering
\includegraphics[scale=0.50]{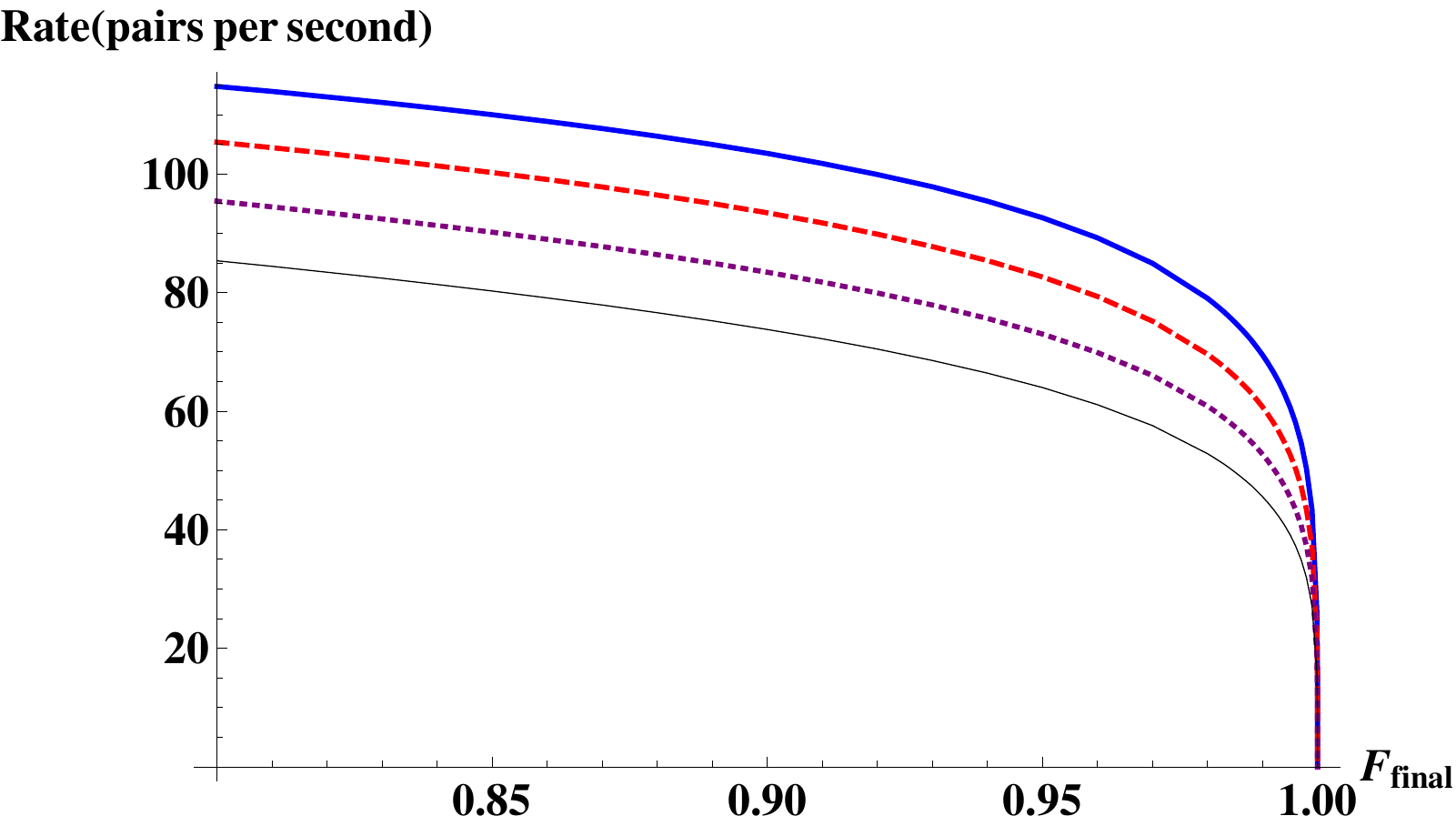}
\caption{(color online). Rates for the hybrid quantum repeater with two rounds of purification for total distance $L= 1280$ km (blue, thick line), $L= 2560$ km (red, dashed line), $L=$ 5120 km (purple, dotted line), $L=$ 10240 km (black, thin line).}
\label{ratesdifdistances}
\end{figure}

Aiming to analyze the length dependence of the rates, in Fig.~(\ref{ratesdifdistances}), we plotted the rates for the hybrid quantum repeater scheme with perfect memories and two rounds of purification in the first nesting level for a total distance of $L=1280$ km, $2560$ km, $5120$ km and $10240$ km. We observe that the rates only decrease inverse-linearly with the total distance. For ideal memories and using purification, we show that it is possible to avoid the exponential decay of the rates with the distance.

\begin{figure}[h!]
\centering
\includegraphics[scale=0.50]{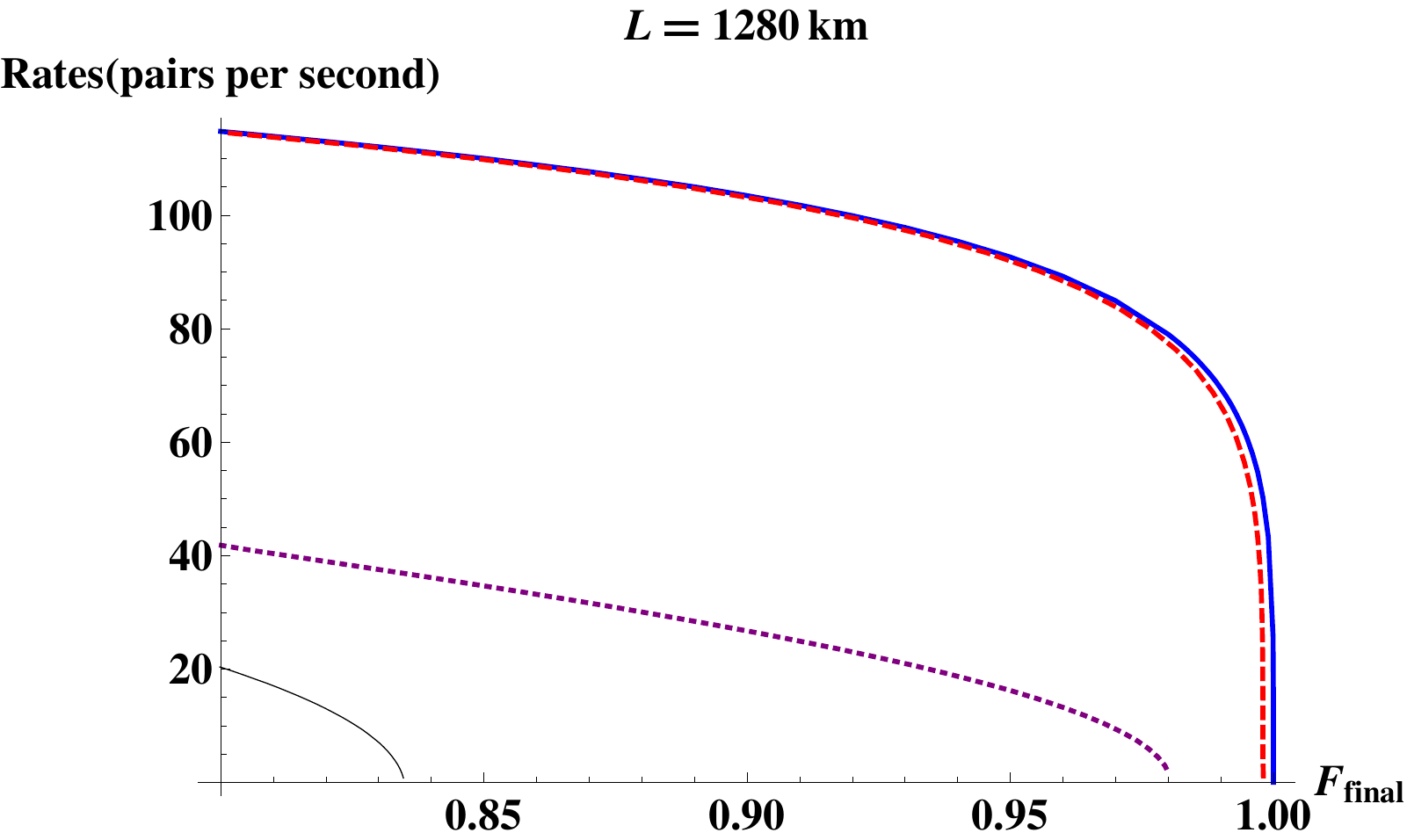}
\caption{(color online). Rates for the hybrid quantum repeater with two rounds of purification for total distance $L= 1280$ km for different parameters of local losses in the controlled-phase gate, $1-T=0$ (blue, thick line), $1-T=0.001\%$ (red, dashed line), $1-T=0.01\%$ (pink, dotted line), $1-T=0.1\%$ (black, thin line).}
\label{l1280loss}
\end{figure}

In Fig.~(\ref{l1280loss}) we can show the rates of the hybrid quantum repeater for $L=1280$ km with different losses in the local gates, T. We show that the protocol can run with local gate errors, however, depending on T, the final fidelity and the generation rates can decrease very fast. For $1-T=0.001\%$ the rates are practically the same as for $T=1$. For $1-T=0.01\%$ and a final fidelity of 0.95, rates of near 16 pairs per second can be achieved. On the other hand, for $1-T=0.1\%$ rates of the order of only 10 pairs per second with a final fidelity below 0.84 can be achieved.

\section{Conclusion}

In this work, we calculated analytically the rates for the so-called
hybrid quantum repeater. Comparing these rates with those obtainable through
direct entanglement distribution (without using swapping, purification, or memories)  
and through quantum relay schemes (using swapping, but no purifications and no memories), 
for sufficiently long distances, the hybrid quantum repeater scheme leads to significantly
better rates. In fact, only the hybrid quantum repeater scales sub-exponentially with the channel length,
while both direct distribution and quantum relay keep the exponential decay with distance. 
We found the rates for a system where the subroutines (generation, swapping, and purification) can all be implemented through weak dispersive light-matter interaction, as proposed in Ref.~\cite{PvLa}.

In the hybrid repeater scheme with homodyne measurement from Ref.~\cite{PvLa}, a rate of 15 pairs per second with final $F=0.98$ was achieved using 16 qubits per half node for a total distance of $L=1280$ km, segment lengths of $L_0=10$ km, and local gate errors of $0.1\%$. In our case, by employing generalized measurements (instead of homodyne projection measurements) for an optimal unambiguous discrimination
of coherent states (USD) as proposed in Ref.~\cite{PvLb}, for just 4 qubits per half node, two rounds of purification in the first nesting level, $L=1280$ km, $L_0=20$ km, and gate errors of $0.001\%$, we achieved rates of the order of 100 pairs per second with $F=0.98$. Note that in the scheme from Ref.~\cite{PvLa} the rates were calculated performing Monte Carlo simulations where not only more rounds of purification in different nesting levels are allowed, but also multiplexing. 
Compared to the banded scheme from Ref.~\cite{vanmeter} for the same distances and final fidelities, we achieve rates with similar order of magnitude, however, it should be pointed out that Ref.~\cite{vanmeter} uses a scheme with initially 50 qubits per half node, and correspondingly more rounds of purification in different nesting levels. 

We achieve our rates (of nearly 100 Hz) in a fairly practical setting where the spatial resources are reduced (e.g. only four cavities per half node),
quantum error detection is close-to-minimal (i.e., only two rounds of purification at the very beginning), and without spatial
multiplexing. However, these results are only obtained under the basic assumptions of
very low local losses, optimal USD measurements for entanglement distribution, deterministic entanglement swapping,
and perfect quantum memories. In future work, we aim to relax one or more of these requirements.
In fact, our analytically obtained rates can be certainly further improved, if we include multiplexing as well as purification in different nesting levels, possibly leading to an improvement of two orders of magnitude compared to the scheme from Ref.~\cite{PvLa}, and similar to the results presented in Ref.~\cite{munro}. Moreover, further improvements in the initial entanglement distribution can be achieved, if instead of coherent states and USD measurements, squeezed states and homodyne measurements (together with extra local phase-space displacements)
are utilized, as was recently shown in Ref.~\cite{ludmila}. Besides its quantitative significance, the work presented here represents a first step
towards a better understanding of the importance of each of the building blocks in the architecture of a hybrid quantum repeater.

\subsection*{Acknowledgments}
Support from the Emmy Noether Program of the Deutsche Forschungsgemeinschaft is gratefully acknowledged. L.P. gratefully acknowledges financial support for the Future and Emerging Technologies (FET) programme within the Seventh Framework Programme for Research of the European Commission, under the FET-Open grant agreement CORNER no. FP7-ICT-213681.


\appendix

\section*{Appendix A}

Purification and entanglement swapping can be performed utilizing controlled-phase (CZ) gates, Hadamard operations, and measurements on the qubits. Assuming that the CZ-gate is the only operation that leads to errors in those schemes, we are able to calculate the final fidelity for the purification and swapping, as well as the probability of purification, using the imperfect CZ-gate obtained in Ref.~\cite{Louis}. These are as follows:

\ba
P_{purification}^{imp}=\frac{1}{2}+ e^{\frac{\pi  (T-1)}{\sqrt{T}}} (F-1) F+\n\\
\left(e^{\frac{\pi
    (T-1) \left(2-i \text{sech}\left(\frac{\log (T)}{2}\right)\right)}{2
   \sqrt{T}}}+e^{\frac{\pi  (T-1) \left(2+i \text{sech}\left(\frac{\log
   (T)}{2}\right)\right)}{2 \sqrt{T}}}\right)\times\n\\
   \frac{(2 (F-1)F+1)}{4},
\ea

\ba
F_{purification}^{imp}=\frac{F+F e^{-\frac{\pi\left(2-2 T\right) }{\sqrt{T}}}}{4P_{purification}^{imp}}+\n\\
\frac{2 F e^{\frac{\pi
   (T-1)}{\sqrt{T}}} \left(\sin \left(\pi
   \left(\frac{3}{2}-\frac{2}{T+1}\right)\right) F+F-1\right)}{4P_{purification}^{imp}},
\ea

\ba
F_{swapping}^{imp}=\frac{1}{4}+\frac{1}{4} e^{2 \pi  \sinh \left(\frac{\log (T)}{2}\right)} (1-2
   F)^2+\n\\
   \frac{e^{\pi  \sinh \left(\frac{\log (T)}{2}\right)}}{2} (2
   (F-1) F+1) \cos \left(\frac{\pi}{2}  \tanh \left(\frac{\log
   (T)}{2}\right)\right).
\ea

\section*{Appendix B}

Consider a coin, which, in one toss, gives ``tail" as an outcome with probability $p$ and ``head" with probability $q=1-p$. Imagine we are just interested in the tail's outcomes. We will here calculate the number of times we have to toss one or more coins to obtain one, two, three or more tails. Note that the problem of calculating the average time necessary to have one success event in this case is analogous to our problem of calculating the average time to successfully generate one (or many) entangled pair(s) in different segments of the repeater.

\paragraph*{\textbf{1) one success}}$\,$\\
\noindent
Imagine that a coin is flipped repeatedly until the first tail
outcome. How many times on average do we have to toss the coin until we get a tail? The average time necessary to obtain one tail is:
\be
\sum_{k=1}^{\infty} k q^{k-1} p =\frac{1}{p} =:\mathcal{Z}_1(p).
\ee

Note that the character $\mathcal{Z}$ is used here in Appendix B for the average time to have tail as an outcome; when average times are calculated to generate entanglement over repeater segments, the character $Z$ is used instead in the main text and Appendix C.

\paragraph*{\textbf{2) two successes}}$\,$\\
\noindent
Now imagine we have two identical coins and we flip both coins at
once until we get a ``tail and tail" outcome. If two tails are
obtained in the same trial, game is over. If ``tail and head"
result is obtained, we keep the tail coin (simulating the situation of memory qubits in the repeater) and flip only the other coin
until the second tail is obtained. In this case, the average time needed to obtain two tails is:
\begin{eqnarray}
\sum_{k=1}^{\infty} k (q^2)^{k-1} p^2+
\sum_{k=1}^{\infty}\sum_{l=1}^{\infty} (k+l) ( (q^2)^{k-1}
2 p q ) (q^{l-1} p )=\nonumber\\
\frac{1}{p}\frac{(3-2p)}{(2-p)}=:\mathcal{Z}_2(p).\,
\end{eqnarray}
The first sum describes the cases when we had the head outcomes in
both coins $(k-1)$ times and then 2 tails were obtained at once.
The second sum counts the cases when at $k$-th trial a tail in one
coin was obtained and stored and then after another $l$ (single)
trials we got the second success.

\paragraph*{\textbf{3) three and more successes}}$\,$\\
\noindent
For 3 successes (in this case 3 tails) in 3 identical coins, the average time needed will be:
\begin{eqnarray}
\!\!\!\!\!\!\sum_{k=1}^{\infty} k (q^3)^{k-1} p^3+\sum_{k=1}^{\infty}
\sum_{l=1}^{\infty} (k+l) ((q^3)^{k-1} 3 p^2 q)  (q^{l-1} p)\nonumber\\+
\sum_{k=1}^{\infty} \sum_{l=1}^{\infty}(k+l) ((q^3)^{k-1} 3 p q^2)
((q^2)^{l-1} p^2)\n\\
\!\!\!\!\!\!+\sum_{k=1}^{\infty} \sum_{l=1}^{\infty} \sum_{m=1}^{\infty}
(k+l+m)( (q^3)^{k-1} 3 p q^2)( (q^2)^{l-1} 2 p q)
 (q^{m-1} p)\n\\=\!\!\frac{p(19+3p(p-4))-11}{p(p-2)(p(p-3)+3)}=:\mathcal{Z}_3(p).\;\;
\end{eqnarray}
In a similar way, we calculate average waiting times for $N$
successes (tails) in $N$ coins and obtain the following recurrence
formula:
\be
\mathcal{Z}_N(p)=\frac{1}{1-q^N}\left(1+
\sum_{j=1}^{N-1} {N \choose j} q^j p^{N-j} \mathcal{Z}_j(p)
\right), \label{rec}
\ee
where
$\mathcal{{Z}}_1(p)=\frac{1}{p}$.

Recurrence formula (\ref{rec}) can be solved, and its solution is given by
\ba
\mathcal{Z}_N(p)=\sum_{k=1}^{N} {N \choose k}
\frac{(-1)^{k+1}}{1-(1-p)^k}\;.
\label{Znapp}
\ea
By substituting $N$ by $2^n$ in Eq.~(\ref{Znapp}), we can calculate the average number of steps necessary to successfully generate entanglement in $2^n$ pairs, as in Eq.~(\ref{Zn}).

\section*{APPENDIX C}


In order to compare the multiplexing with the parallelization strategy, we have to calculate the average times needed to successfully generate entangled pairs over the corresponding segments. We will start our calculation here with the simplest case (two columns and two rows,
similar to Fig.~3 with two memories per half node).


\paragraph*{\textbf{1) two columns ($n=1$), two rows ($r=2$)}}$\,$\\
\noindent
Imagine now that we have 4 coins to toss; coins that can be distinguished
by their positions, arranged into two columns, two rows each. We toss
the coins all at once in the first step, and in the next steps, we toss
only those coins that
had a head outcome in the previous trial.
The average number of steps needed for at least one success (tail) in every column will obviously depend on whether these successes are required to appear in the same row or whether they are allowed to appear in two different rows. The first case
corresponds to parallelization, the second case to multiplexing. 
Probabilities and average numbers of steps are calculated in a
similar manner to Appendix B, taking into account that successes
have to be appropriately distributed among columns and rows. In the following, we will not refer to coins anymore, but to repeater segments and the probabilities to generate entanglement over these segments.

The probability that at least one entangled pair is created in every column in
the case of multiplexing, i.e., when it does not matter in which row the
entanglement was created, is given by:
\ba
&&P_{mult,1,2}=\sum_{k=1}^{\infty} (q^4)^{k-1}
(p^4+4p^3q+4p^2 q^2)\n\\&+&\sum_{k=0}^{\infty} (q^4)^{k-1} 2p^2 q^2
\sum_{l=1}^{\infty} (q^2)^{l-1} (p^2+2pq)\n\\&+&\sum_{k=0}^{\infty}
(q^4)^{k-1} 4p q^3
\sum_{l=1}^{\infty} (q^3)^{l-1} (p^3+3p^2q+2pq^2)\n\\&+&\sum_{k=0}^{\infty}
(q^4)^k 4p q^3
\sum_{l=1}^{\infty} (q^3)^{l-1} pq^2 \sum_{m=1}^{\infty} (q^2)^{m-1}
(p^2+2pq) \;.\n\\\label{Eq7App}
\ea
$P_{mult,n,r}$ is the probability to create at least one entangled pair in every column from the $2^n$ columns with $r$ rows, in the case of multiplexing. The average number of steps needed for at least one success in every column in this case
can be calculated as follows:
\ba
&&Z_{mult,1,2}(q) =\sum_{k=1}^{\infty} (q^4)^{k-1} k
(p^4+4p^3q+4p^2 q^2)\n\\&+&\sum_{k=0}^{\infty} (q^4)^{k-1} 2p^2 q^2
\sum_{l=1}^{\infty} (k+l) (q^2)^{l-1} (p^2+2pq)+\n\\&+&\sum_{k=0}^{\infty}
(q^4)^{k-1} 4p q^3
\sum_{l=1}^{\infty} (k+l) (q^3)^{l-1} (p^3+3p^2q+2pq^2)+\n\\&+&\sum_{k=0}^{\infty}
(q^4)^k 4p q^3
\sum_{l=1}^{\infty} (q^3)^{l-1} pq^2\times\n\\
&&\sum_{m=1}^{\infty}
(q^2)^{m-1} (k+l+m) (p^2+2pq)\n\\&=&\frac{1+2q^2}{1-q^4} \;.\label{Eq8App}
\ea
The probability that at least two ``parallel" entangled pairs are created is:
\ba
&&P_{parallel,1,2}=\sum_{k=1}^{\infty} (q^4)^{k-1}
(p^4+4p^3q+2p^2 q^2)\n\\&+&\sum_{k=0}^{\infty} (q^4)^{k-1} 4p^2 q^2
\sum_{l=1}^{\infty} (q^2)^{l-1} (p^2+2pq)\n\\
&+&\sum_{k=0}^{\infty}
(q^4)^{k-1} 4p q^3
\sum_{l=1}^{\infty} (q^3)^{l-1} (p^3+3p^2q+pq^2)\n\\&+&\sum_{k=0}^{\infty}
(q^4)^k 4p q^3
\sum_{l=1}^{\infty} (q^3)^{l-1} 2pq^2 \sum_{m=1}^{\infty} (q^2)^{m-1}
(p^2+2pq)\,.\n\\
\ea
The corresponding average number of steps needed for parallel successes in two columns ($n=1$) is given
by:
\ba
&&Z_{parallel,1,2}(q)=\sum_{k=1}^{\infty} (q^4)^{k-1} k
(p^4+4p^3q+2p^2 q^2)\n\\&+&\sum_{k=0}^{\infty} (q^4)^{k-1} 4p^2 q^2
\sum_{l=1}^{\infty} (k+l) (q^2)^{l-1} (p^2+2pq)\n\\&+&\sum_{k=0}^{\infty}
(q^4)^{k-1} 4p q^3
\sum_{l=1}^{\infty} (k+l) (q^3)^{l-1} (p^3+3p^2q+pq^2)\n\\&+&\sum_{k=0}^{\infty}
(q^4)^k 4p q^3
\sum_{l=1}^{\infty} (q^3)^{l-1} 2pq^2\times\n\\
&&\sum_{m=1}^{\infty}
(q^2)^{m-1} (k+l+m)
(p^2+2pq)\n\\
&=&\frac{1+q+5q^2+4q^4}{1+q+q^2-q^4-q^5-q^6}\,.
\ea

\paragraph*{\textbf{2) two columns ($n=1$), $r>2$ rows}}$\,$\\
\noindent
Calculations similar to those from Eq.~(\ref{Eq8App}) show that
 for 3, 4,..., r rows (see Fig.~3 with $r$ memories per half node)
 the average number of steps needed for at least one
success in every column in the case of multiplexing is given by:
\ba
 Z_{mult,1,3}(q)=\frac{1+2q^3}{1-q^6},
\ea

\ba
 Z_{mult,1,4}(q)=\frac{1+2q^4}{1-q^8},
\ea
and
\ba
Z_{mult,1,r}(q) =\frac{1+2q^r}{1-q^{2r}},
\label{Zmultapp}
\ea
respectively.

In the main text, Eq.~(\ref{ratemult}) is, except for $T_0$, exactly the inverse of Eq.~(\ref{Zmultapp}) with $q=1-P_0$

\section*{Appendix D}

\begin{figure}[h!]
\centering
\includegraphics[scale=0.50]{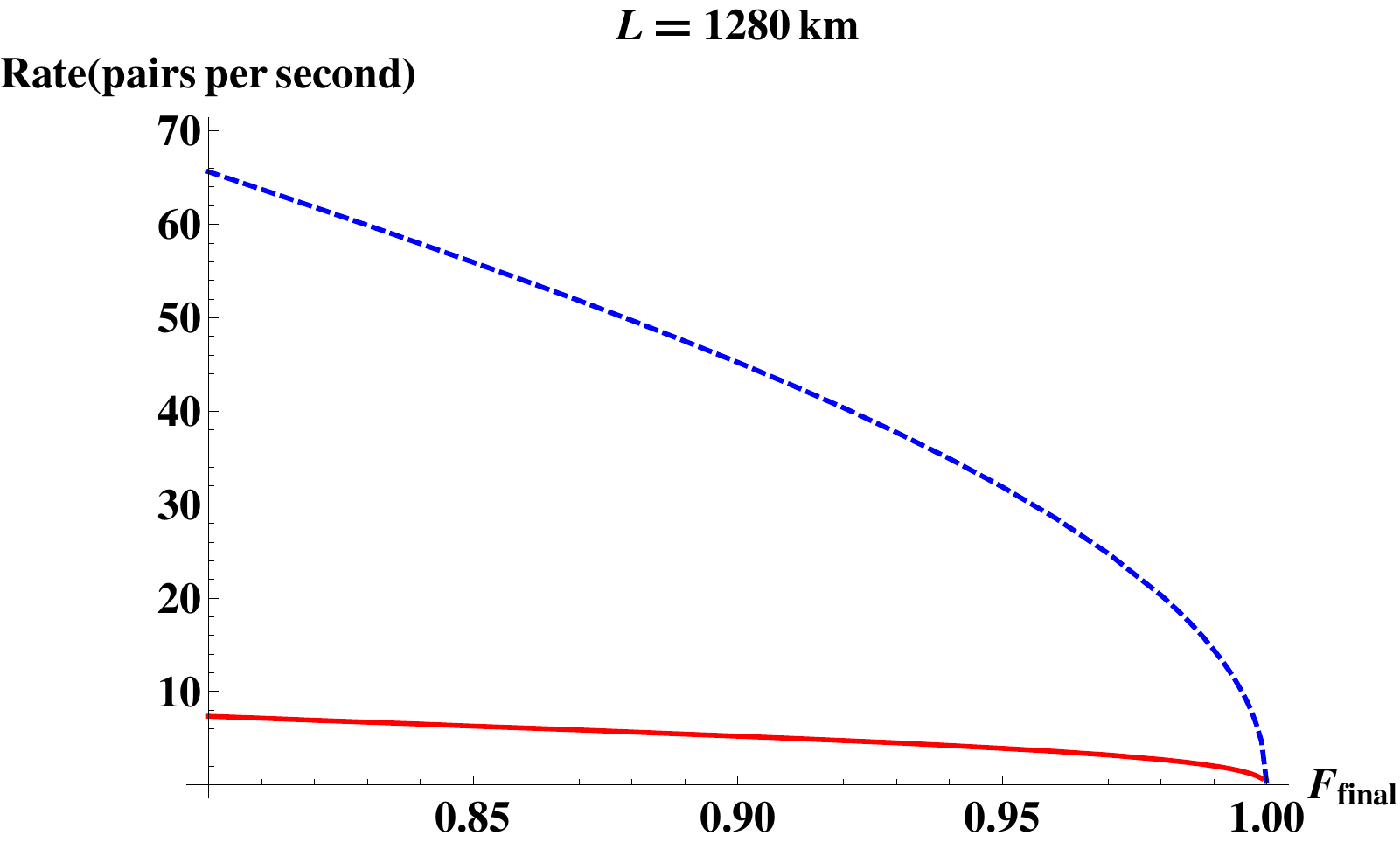}
\caption{(color online). Rates for the hybrid quantum repeater with one round of purification at the first nesting level (blue dashed line) and at the last nesting level (red line) for total distance $L= 1280$ km and $L_0=20$ km.}
\label{ratesbegend}
\end{figure}

The choice of purifying the entangled pairs at the very beginning can be intuitively justified through the probabilistic character of the purification procedure and the deterministic entanglement swapping. Taking as much advantage from the perfect memories as possible, it is reasonable to think that the earlier the purification starts, the smaller the necessary times to generate and purify an entangled pair will be. For a more quantitative justification, let us compare the rates to generate a purified entangled pair in two extreme cases. For purification occurring at the first nesting level, the rate will be calculated as in Eq.~(19). On the other hand, for purification at the end (i.e., at the last nesting level), the rates will be given by $R_{purif,n,end}=\frac{P_1}{T_0Z_{2n}(P_0)}$. Recall that both $P_0$ and $P_1$ are functions of the fidelity and that we are comparing the rates for the same final fidelity. Hence these quantities will have different values in each case. In Fig.~(\ref{ratesbegend}) we illustrate that a scheme where purification occurs in the first level performs better than one with purification at the end.

\end{document}